\documentclass[aps,pra,preprint,showpacs]{revtex4}
\usepackage{amsfonts}
\usepackage{amsmath}

\usepackage{graphicx}

\begin{document}
   
\title{  Wigner Functions and Separability  for Finite Systems}
\author{Arthur O. Pittenger}
\author{Morton H. Rubin} 
\affiliation{Department of Mathematics and Statistics and Department 
of 
Physics\\ 
University of Maryland, Baltimore County,
Baltimore, MD 21250}
\email{rubin@umbc.edu, pittenge@math.umbc.edu}
 \date{\today}
 
 \begin{abstract}
    A discussion of discrete Wigner functions in phase
space related to mutually unbiased bases is presented. This
approach requires mathematical assumptions which limits it to systems 
with
density matrices defined on complex Hilbert spaces of dimension 
$p^{n}$
where $p$ is a prime number. With this limitation it is possible to 
define a phase space and Wigner functions in close analogy to the 
continuous case. 
That is, we use a phase space that is a direct sum of $n$ 
two-dimensional vector spaces each containing $p^{2}$ points. This is 
in contrast to the more usual choice of a two-dimensional phase space 
containing $p^{2n}$ points. A useful aspect of this approach is that
we can relate complete separability of density matrices and their 
Wigner functions in a natural way. 
We discuss this in detail for bipartite systems and present the 
generalization to arbitrary numbers of subsystems when $p$ is odd.
Special attention is required for two qubits $(p=2)$ and our 
technique fails to establish the separability property for more than two qubits.
Finally we give a brief discussion of Hamiltonian dynamics in the 
language developed in the paper. 
\end{abstract}
     
\pacs{03.65.-a, 03.65.Ca, 03.65.Fd}

\maketitle

\section{Introduction}

In a study of thermal equilibrium of quantum systems \cite{Wig32}, 
Wigner introduced the famous function that now bears his name. There is an
extensive literature on the Wigner function for continuous variables 
\cite{review,Folland}. The literature on discrete Wigner functions is less
extensive, but the importance of discrete phase space in quantum 
information has revived interest in the subject \cite{Woottersphase,Vourdas,Paz0}. In
particular, the paper by Gibbons,\ \textit{et. al.} contains a useful 
list of references.

In this paper we present a discussion of discrete Wigner functions in 
phase spaces related to mutually unbiased bases (MUB). Our approach differs 
from the geometric method of Wootters in being more operational and closer 
to the methodology of the continuous case \cite{Woottersphase,Wootters0}, 
but our approach also requires mathematical assumptions which limits it to 
systems with density matrices defined on complex Hilbert spaces of dimension 
$p^{n}$ where $p$ is a prime number. With this limitation it is possible to 
define phase space and Wigner functions which mimic the continuous case. 
There does not seem to be any simple way to do this for other dimensions, see for
example \cite{Leonhardt,Vaccaro}. A useful aspect of this approach is 
that we can relate the separability of density matrices and their Wigner 
functions. 
We discuss this in detail for bipartite systems and present the 
generalization to arbitrary numbers of subsystems. As an application 
of our analysis, we show that for $p$ an odd prime,  with a particular choice of ``phase'' 
parameters, Hermitian operators used in \cite{Woottersphase} for $n$ $p-$level 
systems are tensor products of opeators for the individual $p-$level 
subsystems.

The paper is organized as follows. We first briefly review the 
definition and properties of the Wigner function for continuous variables and 
list the most important properties that are retained in the discrete case. Our
discussion of the discrete Wigner function makes extensive use of
generalized spin matrices which are defined in the section III for a 
singl particle. In order to determine a suitable choice of phase space, we 
are led to consider mutually unbiased bases, and this is done in sections \ref
{MutuallyUB} and \ref{MutuallyUBII}, and further discussed in 
Appendix \ref{appMUB}. The discrete Wigner
function for a single particle is then defined and its properties 
discussed in section \ref{SectionWigner}. The generalization of our discussion 
to more than one particle begins with section \ref{MutuallyUBII}. The 
transition to the general case is aided by using the geometry of discrete phase 
space, which is summarized in Appendix \ref{Geometry}. In section \ref{p=2}
we generalize the Wigner function to dimension $p^{2},$ and in 
\ref{SectWig3} to $p^{n}.$ 

The problem of separability when $p=2$ requires special treatment, 
and in section  \ref{SectWigner2} the case of two qubits is analyzed. The 
generalization to more than two qubits appears to be impossible by 
the present technique, this is discussed in section \ref{SectWig3}. 
Finally, in section IX a brief discussion of Hamiltonian dymanics is 
presented and a simple example using MUB is given.
Various background and technical issues are discussed in the
appendices, including the positivity of the density matrix.

\section{\label{WignerCont}Wigner function for a particle moving in 
one
dimension}

Let $\rho $ be the density matrix for a particle moving in one 
dimension,
and let $Q$ and $P$ be the position and momentum operators for the 
particle.
We set $\hbar =1$ so the Heisenberg commutation relation is $\left[
Q,P\right] =i\mathbf{1}.$ It is convenient to introduce the Wigner 
function
as the Fourier transform of its characteristic function $\chi ,$ 
defined by 
\begin{equation}
\chi_{\rho} (u,v)=tr\left[ \rho D(u,v)\right]  \label{chi}
\end{equation}
where $D$ is the unitary translation operator 
\begin{equation}
D(u,v)=e^{-i(uP-vQ)}=e^{-iuP}e^{ivQ}e^{iuv/2}  \label{D operator}.
\end{equation}
These operators form a projective group called the Heisenberg-Weyl 
group \cite{Weyl}. It is easy to show that 
\begin{equation}
D(u,v)D(a,b)D(u,v)^{\dag }=e^{i(a,b)\circ (u,v)}D(a,b),  
\label{Symplectic D}
\end{equation}
where the phase factor is the symplectic product of the operator
``indices'', 
\begin{equation}
(a,b)\circ (u,v)=bu-av.  \label{sympCont}
\end{equation}
The Wigner function is defined by 
\begin{eqnarray}
W_{\rho}(q,p) &=&\frac{1}{(2\pi )^{2}}\int_{-\infty }^{\infty 
}du\int_{-\infty
}^{\infty }dv\chi_{\rho} (u,v)e^{-i(qv-pu)}  \nonumber \\
&=&\frac{1}{(2\pi )^{2}}\int_{-\infty }^{\infty }du\int_{-\infty 
}^{\infty
}dvtr\left[ \rho D(u,v)\right] e^{-i(qv-pu)}.  \label{Wigner}
\end{eqnarray}

To see that this agrees with the standard definition let us compute 
the
trace in the last equation using a complete set of eigenvectors of 
$Q$, 
\[
W_{\rho}(q,p)=\frac{1}{(2\pi )^{2}}\int du\int dv\int dx\langle 
x|\rho |x+u\rangle
e^{ivx}e^{iuv/2}e^{-i(qv-pu)} 
\]
where Eq.~(\ref{D operator}) was used with 
\[
e^{-iuP}|x\rangle =|x+u\rangle . 
\]
Doing the $v$ and $x$ integrals gives 
\begin{eqnarray}
W_{\rho}е(q,p) &=&\frac{1}{2\pi }\int du\int dx\langle x|\rho 
|x+u\rangle \delta (x+%
\frac{u}{2}-q)e^{ipu}  \nonumber \\
&=&\frac{1}{2\pi }\int du\langle q-u/2|\rho |q+u/2\rangle e^{ipu}.
\label{Wigner2}
\end{eqnarray}
The definition of the operators $D(u,v)$ is not unique. There is some
freedom in the choice of phase, referred to as gauge freedom in 
reference 
\cite{Weyl}, p 181. While the choice used here is the standard one, 
the issue
is not so simple for the discrete case.

Many of the standard properties of the Wigner function can be deduced
readily from Eq.~(\ref{Wigner}):

1. the mapping $\rho \rightarrow W_{\rho}$ is convex linear,

2. $W_{\rho}$ is normalized, \textit{i.e.} 
\[
\int dq\int dpW_{\rho}(q,p)=1 
\]
which follows from $\chi_{\rho}е (0,0)=tr\rho =1,$

3. $W$ is real since $\chi_{\rho}^{*}(u,v)=\chi_{\rho}(-u,-v),$

4 . if $\rho ^{\prime }=D(a,b)^{\dagger }\rho D(a,b)$ then 
\begin{eqnarray*}
W_{\rho\prime}(q,p) &=&\frac{1}{(2\pi )^{2}}\int du\int dvtr\left[ 
\rho
D(a,b)D(u,v)D(a,b)^{\dagger }\right] e^{-i(qv-pu)} \\
&=&\frac{1}{(2\pi )^{2}}\int du\int dvtr\left[ \rho
D(u,v)e^{i(ub-va)}\right] e^{-i(qv-pu)} \\
&=&W_{\rho}(q+a,p+b),
\end{eqnarray*}

5. the marginal distributions are probability densities, 
\begin{eqnarray*}
\int_{-\infty }^{\infty }dqW_{\rho}(q,p) &=&\langle p|\rho |p\rangle 
\\
\int_{-\infty }^{\infty }dpW_{\rho}(q,p) &=&\langle q|\rho |q\rangle .
\end{eqnarray*}
More generally, if we integrate along a line in phase space we get a
probability density 
\[
\int_{-\infty }^{\infty }dq\int_{-\infty }^{\infty 
}dpW_{\rho}(q,p)\delta (q\cos
\theta +p\sin \theta -q_{0}\rangle =\langle q_{0};\theta |\rho 
|q_{0};\theta
\rangle , 
\]
where $|q_{0};\theta \rangle $ is the eigenvector of $Q_{\theta 
}=Q\cos
\theta +P\sin \theta $ with eigenvalue $q_{0}.$

Finally, to show that the Wigner function is equivalent to the density
matrix, we write the density matrix in terms of the Wigner function. 
This is
done easily by taking the inverse Fourier transform of 
Eq.~(\ref{Wigner2}) 
\begin{equation}
\langle q|\rho |q^{\prime }\rangle =\int_{-\infty }^{\infty 
}W_{\rho}(\frac{
q+q^{\prime }}{2},p)e^{-ip(q-q^{\prime })}dp.  \nonumber
\end{equation}
It follows from this equation that 
\[
tr\left[ \rho _{1}\rho _{2}\right] =2\pi \int_{-\infty }^{\infty
}\int_{-\infty }^{\infty }dpdqW_{\rho_{1}}(q,p)W_{\rho_{2}}(q,p), 
\]
which is just Plancheral's theorem.

Proving that a given function $W(q,p)$ corresponds to
a density matrix comes down proving that the inverse formula leads to 
a 
$\rho $ which is positive (cf. ref \cite{Narcowich}).

Finally we note that we can define a Wigner function, $W_{A}$, for 
any 
operator $A$ 
for which Eq.~(\ref{Wigner}) is defined. 

\section{\label{GenerSpinMat}Generalized spin matrices}

We briefly review some facts about the generalized spin matrices 
which are
of interest here and introduce some notation that will be used 
throughout
the paper. We shall use letters $j,k,s,t$ to denote elements of $%
Z_{d}=\{0,1,\cdots ,d-1\},$ the integers modulo $d.$ Let $H_{d}$ be a 
$d$
-dimensional complex Hilbert space, and let $\{|k\rangle ,$ $k\in 
Z_{d}\}$
be an orthonormal basis of $H_{d}.$ Let $M_{d}$ be the vector space of
complex $d\times d$ matrices that act on $H_{d}.$ This space is a 
$d^{2}$
-dimensional Hilbert space with respect to the Frobenius or trace 
inner
product 
\begin{equation}
\langle A,B\rangle =tr\left( A^{\dagger }B\right)  \label{innerprod}
\end{equation}
for $A,B\in M_{d}.$ The set of matrices $\{|j\rangle \langle k|,$ 
$j,k\in
Z_{d}\}$ is an orthonormal basis of $M_{d}.$ Let $\eta =\eta 
_{d}=e^{i2\pi
/d},$ and define the generalized spin matrices as the set of unitary
matrices 
\begin{equation}
S_{j,k}=\sum_{m=0}^{d-1}\eta ^{jm}|m\rangle \langle m+k|  \label{DefS}
\end{equation}
where index addition is to be understood to be modulo $d$. This set 
of $%
d^{2} $ matrices, including the identity matrix $I=S_{0,0}$, forms an
orthogonal basis of $M_{d}$ \cite{PRfourier}.

It is not difficult to show that 
\begin{eqnarray}
S_{j,k}^{\dagger } &=&\eta ^{jk}S_{-j,-k}  \label{HermitianAdj} \\
S_{j,k}S_{s,t} &=&\eta ^{ks}S_{j+s,k+t}.  \label{product}
\end{eqnarray}
From Eq.~(\ref{product}) it follows that $S_{j,k}$ and $S_{s,t}$ 
commute if
and only if the symplectic product $\left( j,k\right) \circ \left(
s,t\right) =0,$ where 
\begin{equation}
\left( j,k\right) \circ \left( s,t\right) \equiv ks-jt\,\bmod{d}
\label{symplectic}
\end{equation}
which should be compared with Eq.~(\ref{sympCont}). We also will need 
the
relation 
\begin{equation}
S_{j,k}^{m}=\eta ^{m(m-1)jk/2}S_{mj,mk},  \label{SPower}
\end{equation}

The spin matrices can be generated from two matrices: $S_{1,0}$ which 
is
diagonal, and $S_{0,1}$ which is real and translates each state to 
the next
lowest one modulo $d$. One can check that 
$S_{j,k}=S_{1,0}^{j}S_{0,1}^{k}.$
These spin matrices can be viewed as translation operators in a manner
similiar to the $D(u,v)$ operators for the single particle discussed 
in
section \ref{WignerCont}.  The analog to property 4 is 
\begin{equation}
S_{s,t}S_{j,k}^{m}S_{s,t}^{\dagger }=\eta ^{m(tj-sk)}S_{j,k}^{m}=\eta
^{m(s,t)\circ (j,k)}S_{j,k}^{m}.  \label{similarity1}
\end{equation}
Since the matrices $\{\frac{1}{\sqrt{d}}S_{j,k}\}$ form an 
orthonormal basis on the $d^{2}$
-dimension Hilbert space $M_{d,}$ they satisfy the completeness 
relation 
\begin{equation}
\frac{1}{d}\sum_{j,k=0}^{d-1}S_{j,k}\text{ }tr(S_{j,k}^{\dagger }A)=A,
\label{completeness}
\end{equation}
where $A\in M_{d}$. This set of spin matrices has appeared repeatedly 
in the
mathematics and physics literature, for example \cite
{Schwinger,WoottersS,Fivel,Ivanovic,PRfourier,Calderbank} among 
others, 
and is often also referred to as the 
(discrete) Heisenberg-Weyl group.

Finally we define a set of orthogonal one-dimensional projection 
operators
that we will need. Let $p$ be a prime number. For $(j,k)\neq (0,0)$ 
and $0\leq r\leq p-1$%
\begin{equation}
P_{j,k}(r)=\frac{1}{p}\sum_{m=0}^{p-1}\left( \alpha_{p}е \left( 
j,k\right) 
\eta
^{r}S_{j,k}\right) ^{m}  \label{Projector}
\end{equation}
where $\alpha_{2}(1,1)=-e^{i\pi/2}$ and $\alpha_{p}е \left(
j,k\right) =1$ otherwise
is a set of orthogonal one dimensional projection operators 
\cite{PRfourier}.
If we make this definition for $d$ not prime, we find that we generate
rank $1$ projection operators which are not orthogonal. The reason 
that the
factor $\alpha_{2}$ appears in the $p=2$ case is that for $p$ an 
odd prime $S^{p}_{j,k}=S_{0,0}$, however, $\left( 
\alpha_{2}(1,1)S_{1,1}\right) ^{2}=S_{0,0}$ since 
$S_{1,1}^{2}=-S_{0,0}.$

\section{\label{MutuallyUB}Mutually unbiased bases I}

We review the theory of mutually unbiased bases (MUB) for a particle 
whose state vectors lie in a $p$-dimensional complex Hilbert space 
$H_{p}$, where $%
p$ is a prime. It can be shown that there exist $p+1$ orthonomal 
bases (ONB)
in this space which are MUB \cite
{Ivanovic,WoottersMUB,PRMUB}; that is, if $\psi $ and $\phi $ are 
state
vectors that belong to a pair of ONB that are mutually unbiased, then 
$|\langle \phi |\psi
\rangle |=1/\sqrt{p}.$ The simplest example of mutually unbiased bases
occurs for $p=2,$ for which the bases are composed of the 
eigenvectors of
the three Pauli matrices $\{\sigma _{x},\sigma _{y},\sigma _{z}\}$.

There is a nice way to characterize the MUB using commuting classes 
of the generalized spin matrices \cite{BBRB}. This leads to a natural 
way to
introduce discrete phase space, and, in turn, to a definition of a 
Wigner
function. We denote the two dimensional vector space with components 
in $%
Z_{p}$ by $V_{2}(p)$, and use the letters $u$ and $v$ to denote 
vectors in
this space. This vector space contains $p^{2}$ distinct points, and 
it is
convenient to index the $p^{2}$е spin matrices using $V_{2}(p),$%
\begin{equation}
v=(v_{0},v_{1})\rightarrow S_{v}=S_{v_{0},v_{1}}.  \label{S mapping}
\end{equation}
With this notation Eq.~(\ref{similarity1}) becomes 
\begin{equation}
S_{u}S_{v}S_{u}^{\dag }=\eta ^{u\circ v}S_{v}.  \label{product1}
\end{equation}
It follows from this that two spin matrices commute if and only if the
symplectic inner product of their \textit{index vectors} vanish. 
Therefore, the
problem of finding commuting sets of operators is transformed into 
finding
solutions to the equation $u\circ v=0$ for vectors in the two 
dimensional
vector space $V_{2}(p).$ The solutions are easy to find; the $p+1$ 
index
vectors $u_{a}$, $a\in I_{p}=\{0,1,\cdots ,p\}$ partition the spin 
matrices
into $p+1$ sets defined by 
\begin{eqnarray}
C_{a} &=&\{bu_{a}=b(1,a),\ b\in Z_{p}\}\rightarrow \mathfrak{M}%
_{a}=\{S_{u_{a}}^{b},\text{ }b\in Z_{p}\}\qquad a<p  \nonumber \\
C_{p} &=&\{bu_{p}=b(0,1),\ b\in Z_{p}\}\rightarrow \mathfrak{M}%
_{p}=\{S_{u_{p}}^{b},\text{ }b\in Z_{p}\}.  \label{MUBs}
\end{eqnarray}
(Note: in \cite{PRMUB} $C_{p}$ was denoted by $C_{\infty }$).

\begin{figure}
\includegraphics[height=3in,width=3.5in]{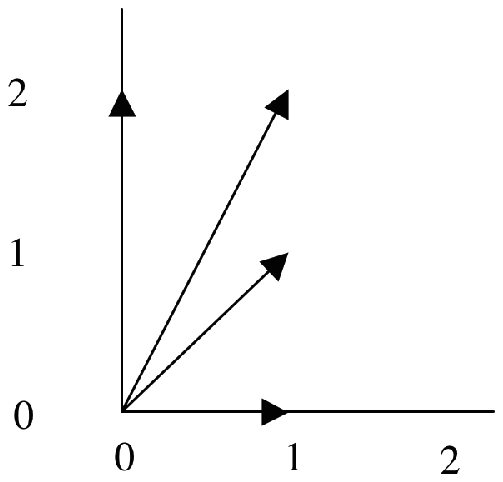}
\caption{\label{fig1} The vectors $u_{a}$ in $V_{2}(3)$.}
\end{figure}

Equation (\ref{MUBs}) relates each vector in $V_{2}(p)$ to commuting 
sets of unitary
matrices such that $\mathfrak{M}_{a}\cap 
\mathfrak{M}_{b}=\{S_{0,0}\}$ for $a\neq b$
. This follows from the fact that in $V_{2}(p)$ two non-zero vectors 
with
vanishing symplectic product must be collinear. The state vectors in 
each
basis are the eigenvectors of the associated set of unitary matrices 
in Eq.~(\ref{MUBs}).
The projection operators for these vectors are defined in 
Eq.~(\ref{Projector}) and can be found in \cite{BBRB}.

$V_{2}(p)$ will be used as the phase space for a single system with 
Hilbert
space $H_{p}$, and vectors in $V_{2}(p)$ will be used as indices for 
the
characteristic function and for  the Wigner function. The 
``horizontal'' 
and ``vertical''
axes of $V_{2}(p)$ are associated with the spin matrices $S_{u_{0}}$ 
and $%
S_{u_{p}}$, respectively. In general, a vector (or point) $(j,k)$ in 
$%
V_{2}(p)$ corresponds to $S_{j,k}.$ The projectors generated by 
$S_{u_{0}}$
are associated with the basis is $\{|j\rangle ,\ j\in Z_{p}\},$ and 
the
projectors generated by $S_{u_{p}}$ are associated with the basis $%
\{|k)=\left( 1/\sqrt{p}\right) \sum_{j=0}^{p-1}\eta ^{kj}|j\rangle ,\ 
k\in
Z_{p}\}.$ The latter states are often referred to as the phase 
states, \cite
{Vourdas,Vaccaro,WoottersS}. The Hermitian operators 
$J=\sum_{j=0}^{p-1}j|j\rangle
\langle j|$ with eigenstates $\{|j\rangle \}$ and $\Phi
=\sum_{k=0}^{p-1}k|k)(k|$ with eigenstates $\{|k)\}$ are said to be
conjugate observables, since these states are Fourier transforms of 
one
another. This is in analogy with the operators $Q$ and $P$ of section 
\ref
{WignerCont} although the commutation relation of $J$ and $\Phi $ is 
not
proportional to the identity operator, and is, therefore, state 
dependent.

The fact that the sets $C_{a}$ correspond to a set of MUB can be seen 
by
computing the projection operators for the sets, and showing that 
\cite{PRMUB}
\begin{eqnarray}
\sum_{r\in Z_{p}}P_{u_{a}}(r) &=&S_{0,0} \\
tr\left[ P_{u_{a}}(r)P_{u_{a}}(s)\right] &=&\delta (r,s)\\
tr\left[ P_{u_{a}}(r)P_{u_{b}}(s)\right] &=&\frac{1}{p}\quad 
\text{for }a\neq
b.  \label{overlap}
\end{eqnarray}
 In particular, the proof of Eq.~(\ref
{overlap}) depends on the orthogonality of the
spin matrices and the fact that $tr(S_{j,k})=0$ for all the spin 
matrices
except the identity. This set of MUB is complete in the sense that
there are $p+1$ ONB in the set, the maximum number possible 
\cite{BBRB}. 

\section{\label{SectionWigner}The discrete Wigner function
for a single particle}

\subsection{The Wigner function}

The discrete Wigner function of interest here was introduced by 
Wootters in 
\cite{Wootters0}. Following Wootters we wish to define the discrete 
analog
of the Wigner function such that properties $1-5$ of section 
\ref{WignerCont}
are preserved. Our approach differs by emphasizing the role of the 
spin
matrices.

Let $p$ be a prime number, and $\rho \in M_{p}$ be a density matrix
describing the state of a system on the Hilbert space $H_{p}.$ Define 
the 
\textit{characteristic function} over $V_{2}(p)$
\begin{equation}
\chi _{\rho }(mu_{a})=\chi (mu_{a})=tr\left[ \rho \left( \alpha_{p}
(u_{a})S_{u_{a}}\right) ^{m}\right],  \label{discrete chi}
\end{equation}
where $\alpha_{p}(u)$ is defined above Eq.~(\ref{Projector}). The
properties of $\chi $ that we shall need are 
\begin{eqnarray}
\chi (0) &=&1  \label{chi0} \\
\chi (mu_{a})^{*} &=&\chi (-mu_{a})  \label{chiCC}
\end{eqnarray}
This last result follows from the fact that $\left( S_{u}^{m}\right)
^{\dagger }=S_{u}^{-m},$ since $S_{u}$ is unitary.

Let $v=(v_{0},v_{1})$ and $u=(u_{0},u_{1})$ be vectors in $V_{2}(p)$. 
Then using Eq.~(\ref{symplectic})
the \textit{discrete Wigner function} is defined as the discrete
symplectic Fourier transform of the characteristic function: 
\begin{eqnarray}
W_{\rho }(v) &=&W(v)=\frac{1}{p^{2}}\sum_{u\in V_{2}(p)}\eta ^{v\circ 
u}\chi
(u)  \nonumber \\
&=&\frac{1}{p^{2}}\left( \chi (0)+\sum_{a=0}^{p}\sum_{m=1}^{p-1}\eta
^{v\circ mu_{a}}\chi (mu_{a})\right) .  \label{discreteWigner}
\end{eqnarray}\textit{}е
The sum over $m$ excludes the $m=0$ term, which gives rise to the 
first term
in brackets. The equality of these two expressions follows from the 
fact
that the vectors $\{mu_{a},$ $a\in I_{p},$ $m\in Z_{p}^{*}\}\cup
\{(0,0)\}=V_{2}(p)$, where $Z_{p}^{*}=Z_{p}-\{0\},$ that is, these 
vectors
partition the space into distinct lines through the origin. This fact
illustrates the role of the geometry of $V_{2}(p)$, see appendix 
\ref{Geometry}.

If we substitute Eq.~(\ref{discrete chi}) into (\ref{discreteWigner}) 
and use 
(\ref{Projector}), the Wigner function can also be written as 
\begin{equation}
W(v)=\frac{1}{p}\left( -1+\sum_{a=0}^{p}tr\left[ \rho P_{u_{a}}\left( 
v\circ
u_{a}\right) \right] \right)  \label{discreteWigner1}
\end{equation}
where $\{pr(r|a,\rho )=Tr\left[ \rho P_{u_{a}}(r)
\right] ,$ $r\in Z_{p}\}$ is the probability distribution that can be
estimated from one of the $p+1$ experiments determined by the set of 
MUB 
\cite{PRMUB}$.$ The sum over $a$ gives a complete set of measurements 
for
determining the Wigner function or, equivalently, as we shall see, the
density matrix. This form of $W$ shows that it is real and that it 
may be negative.

Equation (\ref{discreteWigner1}) can be rewritten as 
\begin{eqnarray}
W(v) &=&tr\left[ \rho A(v)\right]  \nonumber \\
A(v) &=&\frac{1}{p}\left( \sum_{a=0}^{p}P_{u_{a}}\left( v\circ 
u_{a}\right)
-S_{0,0}\right) .  \label{A operator}
\end{eqnarray}
The set of Hermitian operators $\{A(v),$ $v\in V_{2}(p^{2})\}$ was 
used by
Wootters in \cite{Woottersphase} to define the Wigner function and is 
an
orthogonal basis of $M_{p}.$ To verify this, one uses the MUB 
properties
from Eq.~(\ref{overlap}) and computes as in \cite{Woottersphase} 
\[
Tr\left[ A\left( u\right) A\left( v\right) \right] 
=\frac{1}{p^{2}}\left[
p-2\left( p+1\right) +\sum_{a}\sum_{b}Tr\left[ P_{u_{a}}\left( u\circ
u_{a}\right) P_{u_{b}}\left( v\circ u_{b}\right) \right] \right] . 
\]
Regardless of $u$ and $v$ each term in the double sum equals $1/p$ 
when $%
a\neq b.$ If $a=b$ and $u=v$, each of the resulting $p+1$ terms 
equals $1.$
If $a=b$ and $u\neq v$, then the trace equals zero except for the one 
case
when $u-v=mu_{a}$ so that $u\circ u_{a}=v\circ u_{a}$ and the trace 
equals 
one. Collecting terms gives 
\begin{equation}
    Tr\left[ A\left( u\right) A\left( v)\right) \right] 
=\frac{1}{p}\delta \left(
u,v\right).  \label{AsNormalizarion}
\end{equation}
Note, by the way, that one can use the orthogonality to express the 
identity
as 
\begin{equation}
I=\sum_{u}A\left(u\right).  \label{Ieqn}
\end{equation}

In the preceding discussion we have written \textit{the} Wigner 
function 
and \textit{the}
characteristic function. In fact, for a given density matrix and a 
complete
set of MUB, a class of Wigner and characteristic functions can be 
defined.
For example we can multiply the characteristic funtion in 
Eq.~(\ref{discrete
chi}) by an appropriate phase factor and get a new characteristic 
function
\begin{displaymath}
\chi _{\rho }(mu_{a})\rightarrow \chi _{\rho ,r_{a}}(mu_{a})=\eta
^{mr_{a}}\chi _{\rho }(mu_{a}), 
\end{displaymath}
where $r_{a}\in Z_{p}$.
Under this transformation 
\begin{displaymath}
W_{\rho}(v)\rightarrow W_{\rho,\mathbf{r}}(v)=\frac{1}{p}\left( 
-1+\sum_{a=0}^{p}tr\left[
\rho P_{u_{a}}\left( v\circ u_{a}+r_{a}\right) \right] \right) , 
\end{displaymath}
where $\mathbf{r}=(r_{0},\ldots,r_{p})$.
This approach provides an operational way of defining the class of 
Wigner
functions described in \cite{Woottersphase} and in the recent work of 
\cite
{Galvao}.

Before showing that the definition Eq.~(\ref{discreteWigner}) has the
desired properties, we present three examples.

\subsection{Examples}
\subsubsection{Qubits (p=2)}

Using Eq.~(\ref{DefS}), the spin matrices may be shown to be 
equivalent to
the Pauli matrices: 
\[
\left( 
\begin{array}{cc}
S_{0,0} & S_{0,1} \\ 
S_{1,0} & S_{1,1}
\end{array}
\right) =\left( 
\begin{array}{cc}
\sigma _{0} & \sigma _{x} \\ 
\sigma _{z} & i\sigma _{y}
\end{array}
\right) , 
\]
where $\sigma _{0}$ is the $2\times 2$ identity. The classes of MUB 
are
generated by 
\begin{eqnarray*}
C_{0} &=&\{b(1,0)\}\rightarrow \{\sigma _{0},\sigma _{z}\} \\
C_{1} &=&\{b(1,1)\}\rightarrow \{\sigma _{0},i\sigma _{y}\} \\
C_{2} &=&\{b(0,1)\}\rightarrow \{\sigma _{0},\sigma _{x}\}
\end{eqnarray*}
where $b\in Z_{2}$. The most general density matrix may be written as 
\[
\rho =\frac{1}{2}\left( \sigma _{0}+\sum_{j}m_{j}\sigma _{j}\right) 
\]
where $(m_{x},m_{y},m_{z})$ is a vector with real components and 
length less
than or equal to $1.$ In this case
\[
\chi (u_{0})=m_{z},\ \chi (u_{1})=m_{y},\ \chi (u_{2})=m_{x}. 
\]
We have included the factor $\alpha_{2}(u)$ so
that $\chi $ is real. For $p=2$ we have $\eta =-1$, and
for $v=(v_{0},v_{1}) \in V_{2}(2)$
\[
W(v)=\frac{1}{4}\left( 1+m_{z}\eta ^{v_{1}}+m_{y}\eta
^{(v_{1}-v_{0})}+m_{x}\eta ^{-v_{0}}\right) . 
\]
It is now easy to see that summing over a horizontal line gives 
\begin{eqnarray*}
\sum_{v_{0}=0}^{1}W(v) &=&\frac{1}{2}\left( 1+(-1)^{v_{1}}m_{z}\right)
=tr\left[ \rho P_{u_{0}}(v_{1})\right], \\
P_{u_{0}}(0) &=&\frac{1}{2}(\sigma _{0}+\sigma _{z}),\quad 
P_{u_{0}}(1)=\frac{
1}{2}(\sigma _{0}-\sigma _{z}),
\end{eqnarray*}
where $P_{u_{0}}(0)$ is the projection operator for the state 
polarized
along the positive $z$-axis, and $P_{u_{0}}(1)$ is the projection for 
the
state polarized along the negative $z$-axis. A similar result holds 
for the
sum over a vertical line, that is, a sum over $v_{1}$ and the $x$ 
-axis.
For $s\in Z_{2},$%
\begin{eqnarray*}
\sum_{v}W(v)\delta (v\circ u_{2}-s,0) &=&tr\left[ \rho 
P_{u_{1}}(s)\right] \\
&=&\frac{1}{2}\left( 1+(-1)^{s}m_{y}\right)
\end{eqnarray*}
which corresponds to summing along the line $\{b(1,1),$ $b\in 
Z_{2}\}.$
Finally, for this case, the Hermitian matrices defined in Eq.~(\ref{A
operator}) are 
\[
A(v)=\frac{1}{4}\left( \sigma _{0}+\sigma _{z}\eta ^{v_{1}}+\sigma 
_{y}\eta
^{(v_{1}-v_{0})}+\sigma _{x}\eta ^{-v_{0}}\right) . 
\]

It is well-known that for a single particle $W(v)$ can serve as a 
hidden variable probablility
distribution if it is nonnegative. This is because, as we shall see 
below, the measurement of
an arbitrary observable $O$ is given by 
\[
tr(\rho O)=p\sum_{v\in V_{2}(p)}W(v)W_{O}(v), 
\]
where $W_{O}(v)$ is the Wigner function defined with $\rho $ replaced 
by $O$
in Eq.~(\ref{discrete chi}). Therefore, if the Wigner function is 
non-negative we can construct a complete hidden
variable theory of a single qubit consistent with quantum mechanics.
However, there appear to be cases where this does not work, for our 
present
example if $\mathbf{m}=(1,1,1)/\sqrt{3}$, then $W(0,0)<0$ 
\cite{Woottersphase}, see however \cite{Bell} where it is 
shown that a hidden variable theory can always be constructed for a 
single spin. We also note that
since this $\mathbf{m}$ corresponds to a pure state there are bases 
in which 
$W(v)\geq 0.$ The positivity of the Wigner function is therefore  
sufficient but not  necessary for the existence of a hidden 
variable theory. For a discussion of the positivity of the Wigner 
function see 
\cite{Galvao}.

\subsubsection{\label{purestate}A pure state in $H_{p}$ ($p>2$)}

Let $\rho =P_{u_{b}}(r),$ then for 
$a\in I_{p}$%
\begin{eqnarray*}
\chi (mu_{a}) &=&\frac{1}{p}\sum_{k=0}^{p-1}\eta ^{-kr}tr\left(
S_{u_{a}}^{m}\left( S_{u_{b}}^{\dagger}\right) ^{k}\right) \\
&=&\frac{1}{p}\sum_{k=0}^{p-1}\eta ^{-kr}p\delta(a,b)\delta (m,k) \\
&=&\delta (a,b)\eta ^{-rm}.
\end{eqnarray*}

Therefore, 
\begin{eqnarray*}
W\left( v\right) &=&\frac{1}{p^{2}}\left( 1+\sum_{m=1}^{p-1}\eta
^{-m(r+u_{a1}v_{0}-u_{a0}v_{1})}\right) \\
&=&\frac{1}{p}\delta \left( r+u_{a}\circ v,0\right),
\end{eqnarray*}
so that $W(v)$ vanishes except at points along a line in $%
V_{2}(p)$. In particular for the case $a=p,$ $W(v)$ vanishes 
everywhere
except along the vertical line $v_{0}=$ constant, and for $a=0$, 
$W(v)$ is
constant along the horizontal line $v_{1}=$ constant and vanishes 
everywhere
else. 

Given an arbitrary pure state, we can always find a MUB that contains 
this state as one of the basis vectors.   This shows that there is 
always a MUB for which a pure state 
has a non-negative Wigner function. On the other hand if the pure 
state is not chosen as one of the MUB vectors the result is more 
complicated as will be seen in example 4 below.

\subsubsection{\label{Random}Completely random state}

The density matrix for the completely random state is $\rho 
=(1/p)\mathbf{1}%
_{p},$ which gives $W(v)=1/p^{2},$ that is, $W(v)$ is constant. 

\subsubsection{\label{jk operator}The operator $O=|j\rangle \langle 
k|$}

As stated above, we can define a Wigner function for operators other 
than
density matrices. We give an example here which we shall use later. 
For the
case that $p$ is an odd prime, let $|j\rangle $ and $|k\rangle $ be 
vectors in the
standard basis and let 
\[
O=|j\rangle \langle k|. 
\]
Then for $a<p$%
\[
\chi _{j,k}\left( mu_{a}\right) =Tr\left[ \left| j\right\rangle 
\left\langle
k\right| \left( S_{u_{a}}\right) ^{m}\right] , 
\]
and, for reasons that are explained in Section \ref{SectWigner2}, we 
introduce a
phase factor when a=p
\[
\chi _{j,k}\left( mu_{p}\right) =Tr\left[ \left| j\right\rangle 
\left\langle
k\right| \left( \eta ^{-2^{-1}}S_{u_{p}}\right) ^{m}\right] , 
\]
where $-2^{-1}$ is taken as $\left( p-1\right) /2$ since in the 
exponent we can compute $\bmod{p}$. Using Eqs.~(\ref{DefS}) and 
(\ref{SPower}%
), we find 
\begin{eqnarray*}
\chi _{j,k}\left( mu_{a}\right) &=&\eta ^{mk+a\left( m\left( 
m-1\right)
/2\right) }\delta \left( j,ma+k\right) , \\
\chi _{j,k}\left( mu_{p}\right) &=&\eta ^{2^{-1}\left( k-j\right) 
}\delta
\left( j,m+k\right) .
\end{eqnarray*}
Working through the details gives \begin{equation}
W_{\left| j\right\rangle \left\langle k\right| }\left( v\right) 
=\frac{1}{p}%
\eta ^{\left( v_{0}+2^{-1}\right) \left( k-j\right) }\delta \left(
0,v_{1}+2^{-1}\left( j+k\right) \right) ,  \label{pointwig}
\end{equation}
where $v=(v_{0},v_{1}).$ Note that if $k=j$, $|j\rangle
\langle j|$ is a density and $W_{| j\rangle
\langle j|}$ is a special case of example 2 above. For $j\neq k,$ we 
get 
\[
W_{O}(v)^{*}=W_{O^{\dagger }}(v). 
\]

Now suppose that $|\psi\rangle=\sum_{j=0}^{p-1}c_{j}|j\rangle$, then 
\[ 
W_{|\psi\rangle \langle \psi|}(v)=\frac{1}{p}\sum_{r=0}^{p-1} 
\eta^{(v_{0}+2^{-1})r}c_{-v_{1}-2^{-1}r}c_{-v_{1}+2^{-1}r}^{*}.
\]
As stated above this is a more complicated form than we found for the 
case $|\psi\rangle \langle \psi|=P_{b}(r).$ For the case $p=3$ we 
illustrate this in Fig. 2 for the case $|\psi\rangle 
=\frac{1}{\sqrt{2}}(|1\rangle+|2\rangle).$

\begin{figure}
\includegraphics[height=4in]{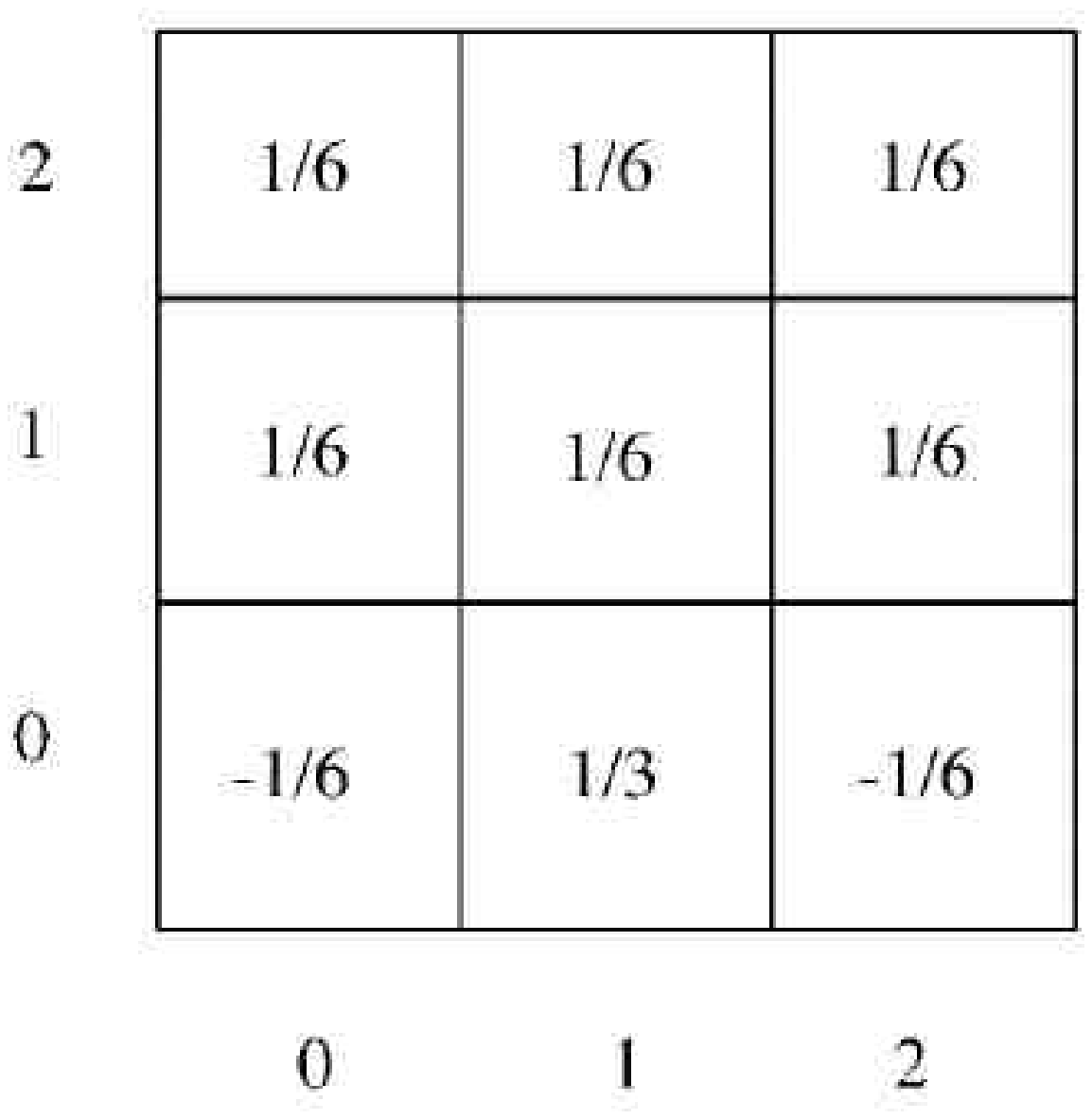}
\caption{\label{fig2} The Wigner function for $p=3$ for the pure 
state $|\psi\rangle =\frac{1}{\sqrt{2}}(|1\rangle+|2\rangle)$.}
\end{figure}

\subsection{\label{Properties}Properties of the Discrete
Wigner Function}

We now examine whether the definition (\ref{discreteWigner}) or,
equivalently, (\ref{discreteWigner1}) satisfies the criteria that we 
set out
in part 1.

1. The mapping is $\rho \rightarrow W_{\rho}$ is linear on $M_{d}$ 
and 
convex linear on the density matrices.

2. $W(v)$ is normalized since 
\[
\sum_{v_{0}v_{1}=0}^{p-1}W(v)=\frac{1}{p^{2}}\left(
p^{2}+\sum_{a=0}^{p}\sum_{m=1}^{p-1}\chi (mu_{a})p^{2}\delta 
(m,0)\right)
=1. 
\]

3. The reality of $W$ follows immediately from 
Eq.~(\ref{discreteWigner1}).

The first three results also follow directly from Eqs.~(\ref{A 
operator}) and
(\ref{Ieqn}).

4. For $w\in V_{2}(p)$, if $\rho ^{\prime }=S_{w}^{\dagger }\rho 
S_{w}$ then using Eq.~(\ref{product1}) 
\begin{eqnarray*}
\chi _{\rho ^{\prime }}(u) &=&\eta ^{w\circ u}\chi _{\rho }(u) \\
W_{\rho ^{\prime }}(v) &=&W_{\rho }(v+w).
\end{eqnarray*}
Note that if $\rho$ commutes with $S_{w}$, the Wigner function is 
invariant under translations along $w$. Furthermore, the 
characteristic function vanishes for $u$ such that $w\circ u \neq 0 
\bmod 
p$.

5. The marginal distributions are easily computed. We consider the 
more
general case of summing along the points on any of the lines in phase 
space,
where a line in phase space $V_{2}(p)$ is defined as the set of 
points that
satisfy the equation 
\begin{eqnarray}
L(b,s) &=&\{(x,y):-xb+y-s=(x,y)\circ u_{b}-s=0,\quad b,s\in Z_{p}\} 
\nonumber \\
L(p,s) &=&\{(s,y):-x+s=(x,y)\circ u_{p}+s=0,\quad b,s\in Z_{p}\}. 
\label{line}
\end{eqnarray}
$L(b,s)$ is the line with ``slope'' $b$\ which intersects the 
vertical axis at 
$s,$ and $L(p,s)$ is a ``vertical'' line that intersects the 
horizontal axis at 
$x=s$ (see Appendix \ref{Geometry}). Let 
\[
f_{b}(s)=\sum_{v}W(v)\delta (v\circ u_{b}-s,0), 
\]
then using Eqs.~(\ref{discreteWigner1}) and (\ref{overlap}), we can 
show
that 
\begin{eqnarray*}
f_{b}(s) &=&\frac{1}{p}\left( -p+\sum_{a\neq b}tr\left[ \rho
\sum_{r=0}^{p-1}P_{u_{a}}(r)\right] +p\ tr\left[ \rho 
P_{u_{b}}(s)\right]
\right) \\
&=&tr\left[ \rho P_{u_{b}}(s)\right] .
\end{eqnarray*}
We have used the fact that for $a\neq b$\ the sum over $v$\ becomes a 
sum
over $Z_{p}$,\ and this sum is the identity operator, while for 
$a=b$, 
we have $v\circ u_{b}=s.$\ Therefore, we see that summing the Wigner 
function over
any line in phase space gives the probablilty that the system is in 
the
corresponding MUB state.

6. Since $W_{\rho}$ and $\chi_{\rho} $ are Fourier transforms of one 
another, Plancheral's
formula gives 
\begin{equation}
p^{2}\sum_{v}W_{\rho}(v)^{2}=|\chi_{\rho} 
(0)|^{2}+\sum_{a=0}^{p}\sum_{m=1}^{p-1}|\chi_{\rho}
(mu_{a})|^{2}.  \label{Pancheral}
\end{equation}
We also have, setting $mu_{a}=(j,k)=v,$
\[
|\chi_{\rho}е(v)|^{2}=|tr\rho S_{v}|^{2}=\langle \rho ,S_{v}\rangle 
\langle
S_{v},\rho \rangle . 
\]
using Eq.~(\ref{innerprod}). More generally,
\[
\chi _{\rho_{1}}^{*}(v)\chi _{\rho_{2}}(v)=\left( tr\rho 
_{1}S_{v}\right) ^{*}tr\rho
_{2}S_{v}=\langle \rho _{1},S_{v}\rangle \langle S_{v},\rho 
_{2}\rangle.
\]
Summing over the complete set of $S_{v}$, from 
Eq.~(\ref{completeness}), we can write Plancheral's formula, 
\begin{equation}
tr\left[ \rho _{1}\rho _{2}\right] 
=p\sum_{v}W_{\rho_{1}}(v)W_{\rho_{2}}(v).
\label{quadratic}
\end{equation}
See also \cite{Woottersphase} where the derivation is based on 
Eq.~(\ref{AsNormalizarion}).

The support of a function $f(v)$ on phase space is defined by 
\begin{equation}
supp(f)=\{v\in V_{2}(p):\text{ }f(v)\neq 0\}  \label{supp}
\end{equation}
and $|supp(f)|$ is defined as the number of points in $supp(f).$ From 
Eq.~(\ref
{quadratic}) we have 
\begin{equation}
pW_{\rho}^{2}(v_{0})\leq p\sum_{v}W_{\rho}^{2}(v)=tr\rho ^{2}\leq 1.  
\label{supp1}
\end{equation}
which implies that for any point $v_{0},$%
\[
|W(v_{0})|\leq \frac{1}{\sqrt{p}}. 
\]

Now let 
\[
\mu _{W_{\rho }}(v)=\left\{ 
\begin{array}{ll}
1 & v\in suppW_{\rho } \\ 
0 & \text{otherwise.}
\end{array}
\right. 
\]
Then applying the Schwarz inequality to the normalization equation 
and 
using Eq.~(\ref{supp1}) we get 
\[
1=\sum_{v}W_{\rho }(v)\mu _{W_{\rho }}(v)\leq \sqrt{\sum_{v}W_{\rho 
}^{{2}}е(v)\mu _{W_{\rho }}(v)\sum_{u}\mu_{W_{\rho}е}е(u)} \leq 
\sqrt{\frac{1}{p}|suppW_{\rho }|}
\]
or 
\[
|suppW_{\rho}|\geq p. 
\]
This is analogous to the continuous
case where the uncertainty principle implies that $W(q,p)$ can not be
concentrated into too small a region. 
We have seen that if $\rho $ is a pure 
state selected from the MUB that $|suppW_{\rho}|=p,$ and $%
W_{\rho }(v)=1/p$ on its support, so the lower bound is attained.

If $\rho _{1}$ and $\rho _{2}$ correspond to orthogonal states, then 
Eq.~(\ref
{quadratic}) gives 
\[
\sum_{v}W_{\rho_{1}е}(v)W_{\rho_{2}е}(v)=0, 
\]
which along with the normalization condition implies that, either 
$suppW_{\rho_{1}е}$
and $suppW_{\rho_{2}}$ are disjoint or at least one of the Wigner 
functions must
take on negative values. For example, we saw in \ref{purestate} that 
the 
orthogonal states in one of the bases of a set of MUB have support on 
non-intersecting lines of $V_{2}(p)$.  

There is an inequality, referred to as an uncertainty principle, that 
also
follows from the discrete Fourier transform: 
\[
|suppW_{\rho}е|\text{ }|supp\chi_{\rho}е |\geq p^{2}, 
\]
\cite{Terras}. Equality holds for the random state discussed in 
example \ref
{Random} above.

\subsection{Inversion formula}

In the case of continuous phase space, the density matrix for a 
particle
confined to one-dimension can be obtained from Eq.~(\ref{Wigner2}) by 
using
the inverse Fourier integral. We can proceed in a similiar manner for 
the
discrete case. First using the discrete Fourier inversion formula, 
\begin{equation}
\chi \left( mu_{a}\right) =\sum_{v\in V(p^{2})}W(v)\eta ^{-v\circ 
mu_{a}}
\label{inverseFT}
\end{equation}
Then from Eq.~(\ref{discrete chi}) and the completeness of the spin 
matrices 
\begin{equation}
\rho =\frac{1}{p}\left( S_{0,0}+\sum_{a=0}^{p}\sum_{m=1}^{p-1}\chi 
\left(
mu_{a}\right) ^{*}S_{u_{a}}^{m}\right) .  \label{inversion}
\end{equation}
Substituting (\ref{inverseFT}) into (\ref{inversion}), and using Eqs. 
(\ref
{Projector}) and (\ref{A operator}), we also get 
\begin{equation}
\rho =\sum_{v\in V(p^{2})}W(v)A(v).  \label{inversion1}
\end{equation}
Therefore, we have an expression for the density matrix as an 
expansion in
the spin matrices with coefficients given by the characteristic 
function
and an equivalent expansion in terms of a basis of Hermitian 
operators with the Wigner
function as coefficients.

\section{\label{MutuallyUBII}Mutually Unbiased Bases II}

To define the generalized spin matrices in the case when $d=p^{n}$ 
where $p$
is prime, we require the notion of a finite or Galois field 
$GF(p^{n}),$ see
Appendices \ref{FiniteFields} and \ref{appMUB} for more details. 
There is a
systematic way of representing the elements in $GF\left( p^{n}\right) 
$ that
uses the structure of polynomials irreducible over $GF\left( p\right)
=Z_{p}. $ An irreducible polynomial is a polynomial $f(x)$ of degree 
$n$
with coefficients in $GF(p)$ that can not be factored into 
non-constant polynomials of
lower degree. Then the elements of $GF(p^{n})$ may be represented by 
polynomials of degree less than $n$ with coefficients in $GF(p)$.
The simplest example is that of two
qubits, $p=2,$ $n=2$. In this case the irreducible polynomial is 
unique 
and is
given by $x^{2}+x+1$. Define $\lambda $ to be a symbolic solution of 
$%
x^{2}+x+1=0$ $\bmod{2}.$ Then every element of $GF(2^{2})$ can be
written as $\alpha =a_{0}+\lambda a_{1}$ where $a_{0}$ and $a_{1}$ 
are in 
$GF(2)$ . This is analogous when working with real numbers to letting 
$i$ denote a symbolic solution of
the equation $x^{2}+1=0$ and introducing complex numbers as $x+iy.$ 

For the case of $n=2$ and $p$ an odd prime, let $D$ be an element in 
$%
GF\left( p^{2}\right) $ such that there is no solution in 
$Z_{p}=GF(p)$ to
the equation $x^{2}-D=0$ $\bmod{p}$. In technical terms, $D$ is a
\textit{quadratic non-residue} of $p$. There are an equal number of 
\textit{quadratic residues} and quadratic non-residues in $GF(p)$. 
Then elements in $GF\left(
p^{2}\right) $ can be represented as $j+k\lambda $, where $j$ and $k$ 
are in 
$GF\left( p\right) $ and $\lambda $ is taken to be a symbolic 
solution of $%
x^{2}-D=0$ $\bmod{p}$ . Addition and multiplication of elements of 
$GF(p^{2})$ are defined by 
\begin{eqnarray*}
  (j_{1}+k_{1}\lambda )+( j_{2}+k_{2}\lambda) 
  &=&(j_{1}+j_{2})+(k_{1}+k_{2})\lambda \\ 
(j_{1}+k_{1}\lambda) ( j_{2}+k_{2}\lambda ) &=&(
j_{1}j_{2}+Dk_{1}k_{2}) +( j_{1}k_{2}+k_{1}j_{2}) \lambda , 
\end{eqnarray*}
where the additions in the parentheses are modulo $p$. We refer to 
Appendix 
\ref{appMUB} for more details.

We can construct a complete set of mutually unbiased bases when 
$d=p^{n}$ by following the same procedure that
was used in the $d=p$ case \cite{Wootters0}е. The key idea for 
constructing a MUB is based on
the fact that we can define a two-dimensional vector space 
$V_{2}(p^{n})$
over $GF(p^{n})$, and $p^{n}+1$ generating vectors $u_{\alpha }$ 
where $%
\alpha $ is in the index set $I_{p^{n}}=GF(p^{n})\cup \{p^{n}\}.$
Specifically, define 
\begin{equation}    
u_{\alpha }=\left\{ 
\begin{array}{ll}
(1,\alpha ), & \alpha \in GF(p^{n}) \\ 
(0,1), & \alpha =p^{n}
\end{array}
\right.    \label{ueq}
\end{equation}
Each of these vectors can be used to define a class containing $p^{n}$
vectors, 
\begin{equation}
C_{\alpha }=\{\beta u_{\alpha },\beta \in GF(p^{n})\}
\label{Calpha}
\end{equation}
where $\alpha \in I_{p^{n}}.$ Each pair of vectors in a class has 
vanishing
symplectic product, Eq.~{}(\ref{symplectic}) where the operations are 
with respect to $GF(p^{n})$. We
want to find a spin matrix representation of these classes, that is, 
we wish
to find a mapping from this space to the set of tensor products 
\begin{equation}
S_{\mathbf{u}}=\bigotimes_{r=0}^{n-1}S_{u_{r}}  \label{TensProd}
\end{equation}
where $\mathbf{u}=\bigoplus\limits_{r=0}^{n-1}u^{(r)}$ $\in 
V_{2n}(p)$ and
each $u^{(r)}е\in V_{2}^{(r)}(p)=V_{2}(p).$ To do this we define an 
isomorphism 
\[
M:V_{2}(p^{n})\rightarrow 
V_{2n}(p)=\bigoplus_{j=0}^{n-1}V_{2}^{(j)}(p) 
\]
that preserves the symplectic product in the following sense. For 
each vector 
$v\in V_{2}(p^{n}),$ if  
$M(v)=\mathbf{v}=\bigoplus_{j=0}^{n-1}u^{(j)},$ where
 $u^{(j)}\epsilon V_{2}^{(j)}(p),$ define 
\begin{equation}
\mathbf{v}_{1}\circ \mathbf{v}_{2}=\sum_{j=0}^{p-1}(u_{1}^{(j)}\circ
u_{2}^{(j)}).  \label{syprmap}
\end{equation}
Then $v_{1}\circ v_{2}=0$ implies $\mathbf{v}_{1}\circ 
\mathbf{v}_{2}=0$ as is shown in Eq.~(\ref{sympprodmap}).
We present an outline of the derivation of $M$ in Appendix 
\ref{appMUB} and
refer to \cite{PRMUB} for another discussion. It is worth noting that 
the 
construction of $V_{2n}(p)$ is analogous to what is done in the 
continuous case. There we take the direct sum of the two-dimensional 
vector spaces corresponding to independent conjugate position and 
momentum 
pairs.

To perform the analog of what was done in Eq.~(\ref{MUBs}), it is 
useful to introduce the generators of the index set for
the set of MUB, again the details are given in Appendix \ref{appMUB}. 
For $%
\alpha \in I_{p^{n}},$ define $\lambda ^{r}u_{\alpha }\in 
V_{2}(p^{n}),\
r=0,\cdots ,n-1.$ Then define a set of generators on $V_{2n}(p)$ 
\begin{equation}
G_{\alpha }=\{\mathbf{g}_{r}(\alpha )=M\left( \lambda ^{r}u_{\alpha 
}\right)
,\text{ }r=0,\cdots ,n-1\},  \label{Generators}
\end{equation}
and define the corresponding spin matrix using Eq.~(\ref{TensProd}) 
as 
\begin{equation}
S_{\mathbf{g}_{r}(\alpha )}^{b}\equiv
\bigotimes_{j=0}^{n-1}S_{u_{r}^{(j)}(\alpha)}^{b}  \label{Smatrixpp}
\end{equation}
where each $S_{u^{(j)}}$ acts on a Hilbert space $H_{p}.$  The
generalization of Eq.~(\ref{MUBs}) is
\begin{equation}
G_{\alpha }\rightarrow \mathfrak{M}_{\alpha}=\left\{ 
\prod\limits_{r=0}^{n-1}S_{\mathbf{g}%
_{r}(\alpha
)}^{b_{r}}=\bigotimes\limits_{j=0}^{n-1}\prod%
\limits_{r=0}^{n-1}S_{u_{r}^{(j)}(\alpha )}^{b_{r}},\quad b_{r}\in
GF(p)\right\}  \label{MUBsn}
\end{equation}
for the generation of $p^{n}+1$ disjoint sets of $p^{n}$ of commuting
operators $\mathfrak{M}_{\alpha }$ where $\mathfrak{M}_{\alpha }\cap 
\mathfrak{M}_{\beta
}=\{S_{0,0}\}$ for all $\alpha \neq \beta.$ We have written the 
mapping in Eq.~(\ref{MUBsn}) from the set of basis vectors 
$G_{\alpha}$ 
rather than the space $C_{\alpha}$.

It is also possible to write down the set $\left\{ P_{\alpha }\left(
r\right) ,r\in V_{n}\left( p\right) \right\} $ of rank one orthogonal
projections defined by each of the $p^{n}+1$ commuting classes 
$\mathfrak{M}%
_{\alpha }.$ This gives the set of MUB as projections defined 
explicitly in
terms of sums of the spin matrices in each class. The procedure to do 
this
is discussed in \cite{PRMUB}, and is illustrated there for the case 
for $n=2$
. The corresponding projection operators for the case $p>2$ are 
\begin{eqnarray}
P_{\alpha }(\mathbf{s})&=&\prod_{r=0}^{n-1} P_{\mathbf{g}_{r}(\alpha 
)}(s_{r})
\nonumber \\
P_{\mathbf{g}_{r}(\alpha )}(s_{r})&=&\left(\frac{1}{p}
\sum_{b_{r}=0}^{p-1}\left[ \eta ^{s_{r}}S_{\mathbf{g}_{r}(\alpha 
)}\right]
^{b_{r}}\right) ,  \label{Projector2}
\end{eqnarray}
where $\mathbf{s}=(s_{0},\ldots ,s_{n-1}).$ For $p=2$ it is necessary 
to
include the factors $\alpha_{2}(j,k)$ in the definition of the 
projection
operators as shown in Eq.~(\ref{Projector}).
 $P_{\alpha }\left( \mathbf{r}\right) $ has trace one, and it is
straightforward to check that if $\mathbf{r}\neq \mathbf{s}$ 
\[
P_{\alpha}(\mathbf{r})P_{\alpha}(\mathbf{s})=\delta(\mathbf{r},\mathbf{s})P_{\alpha}(\mathbf{r}).
\]
It is easy to show that each $P_{\alpha }\left(
\mathbf{r}\right) $ is a product of commuting projections. One can 
also show that 
$P_{\alpha }\left( \mathbf{r}\right) =\left( P_{\alpha }\left( 
\mathbf{r}\right) \right)
^{\dagger }$, and it follows that $P_{\alpha 
}\left(\mathbf{r}\right)$ is a rank
one orthogonal projection and that 
\[
I=\sum_{\mathbf{r}}P_{\alpha }\left(\mathbf{r}\right) . 
\]
Finally, it can be shown that for $\alpha\neq\beta$ that
\[
tr\left[P_{\alpha}(\mathbf{r})P_{\beta}(\mathbf{s})\right]е=\frac{1}{1/p^{n}}.
\]

The explicit calculation of the projections and of the set of vectors 
in $%
V_{2n}\left( p\right) $ corresponding to $C_{\alpha }$ depends quite
specifically on $p$ and $n$ and on the representation of elements in 
the
different finite fields. When $n=2$ and $p$ is an odd prime, however, 
one
can give a unified summary of the results of the theory. Without going
through the detailed construction outlined in Appendix \ref{appMUB}, 
it is
easy to check that the vectors in each of the classes below have 
symplectic product
zero.

\noindent \textit{Example d=p$^{2}$}

Let $d=p^{2}$ with $p$ an odd prime and $D$ such that $x^{2}-D=0$ 
$\bmod{p}$ has no solution in $GF(p).$ In Appendix \ref{appMUB}, 
Eq.~(\ref
{Genpsquared}) it is shown that the $p^{2}+1$ commuting classes of 
indices
are generated by 
\[
G_{a_{0},a_{1}}=\left\{ \left( 1,2a_{0},0,2Da_{1}\right) ,\left(
0,2Da_{1},1,2Da_{0})\right) \right\} 
\]
where $a_{0}$ and $a_{1}$ are in $GF(p)$, and 
\[
G_{p^{2}}=\left\{ \left( 0,1,0,0\right) ,\left( 0,0,0,1)\right) 
\right\} . 
\]
One can check directly that the vectors in each $G_{a_{0},a_{1}}$ 
have vanishing symplectic product.
Then the spin matrices that generate the commuting classes may be 
written as 
\begin{eqnarray}
G_{a_{0},a_{1}} &\rightarrow &\mathfrak{M}_{(a_{0},a_{1})}=\left\{ 
\left(
S_{1,2a_{0}}\otimes S_{0,2Da_{1}}\right) ^{b_{0}}\left( 
S_{0,2Da_{1}}\otimes
S_{1,2Da_{0}}\right) ^{b_{1}},\text{ }b_{0},b_{1}\in GF(p)\right\} , 
\nonumber \\
G_{p^{2}} &\rightarrow &\mathfrak{M}_{p^{2}}=\left\{ \left(
S_{0,1}^{b_{0}}\otimes S_{0,0}\right) \left( S_{0,0}\otimes
S_{0,1}^{b_{1}}\right) ,\ b_{0},b_{1}\in GF(p)\right\} .
\label{SpinMatpsquar}
\end{eqnarray}
The corresponding projections are given by 
\begin{eqnarray*}
P_{a_{0},a_{1}}\left( \left( r_{0},r_{1}\right) \right) 
&=&\frac{1}{p}%
\sum_{b_{0}=0}^{p-1}\left( \left( \eta ^{r_{0}}S_{1,2a_{0}}\otimes
S_{0,2Da_{1}}\right) ^{b_{0}}\right) 
\frac{1}{p}\sum_{b_{1}=0}^{p-1}\left(
\left( \eta ^{r_{1}}S_{0,2Da_{1}}\otimes S_{1,2Da_{0}}\right) 
^{b_{1}}\right)
\\
P_{p^{2}}\left( \left( r_{0},r_{1}\right) \right) &=&\frac{1}{p}
\sum_{b_{0}=0}^{p-1}\left( \left( \eta ^{r_{0}}S_{0,1}\otimes 
S_{0,0}\right)
^{b_{0}}\right) \frac{1}{p}\sum_{b_{1}=0}^{p-1}\left( \left( \eta
^{r_{1}}S_{0,0}\otimes S_{0,1}\right) ^{b_{1}}\right) .
\end{eqnarray*}
We note that each of these one-dimensional projection operators is the
product of two commuting rank $p$-dimenional projections. The two $p$
-dimensional spaces that they project onto intersect in a
one-dimensional space.

\section{\label{SectWigner2}Wigner Function For $d=p^{2}$}

In earlier work \cite{Woottersphase,Paz}, the phase space on which 
the Wigner functions 
were defined when $d = p^{n}$ was chosen to be $V_{2}(p^{n})$. The 
advantage of this 
choice is that one can use the underlying geometry to great 
advantage. 
The disadvantage is that one has to label coordinates using elements 
from the 
Galois field $GF(p^{n})$ which does not lend itself to a discussion 
of separability. However, as we saw in Section \ref{MutuallyUBII}, 
and as is elaborated in Appendix \ref{appMUB}, there is a natural 
isomorphism 
M between $V_{2}(p^{n})$ 
and $V_{2n}(p)$ which encodes the geometry of $V_{2}(p^{n})$ in 
$V_{2n}(p)$. We take 
advantage of this structure to define our Wigner function on 
$V_{2n}(p)$  This is in close analogy to the continuous case and 
simplifies
computations involving the generalized 
spin matrices.

In particular, this approach enables questions involving separability 
to be 
treated efficiently. In this section we illustrate the ideas in 
detail for $n = 2$, leaving the generalizations to the next section 
and the Appendix.

\subsection{\textbf{Separability of the Wigner Function for $p$ an 
odd 
prime}}

We consider a bipartite system composed of subsystems of dimension 
$p,$ a
prime.  As we saw in Section \ref{Wigner}, there is a certain 
latitude in
the definition of the Wigner function that is available because of the
freedom to include phase factors in the characteristic function. Our 
goal in
this section is to show how that freedom enables us to define Wigner
functions for one and two subsystems so that separability is 
respected.
Specifically, for a product state we want 
\begin{equation}
\rho =\tau \otimes \mu \Rightarrow W_{\rho }\left( \mathbf{u}\right)
=W_{\tau }\left( u^{(0)}\right) W_{\mu }\left( u^{(1)}\right) ,
\label{wigsep2}
\end{equation}
where $\mathbf{u}=u^{(0)}\oplus u^{(1)}.$ Then, since $W_{\rho 
}\left( 
\mathbf{u}\right) $ is convex linear on the space of densities, we 
will have the
general statement that 
\begin{equation}
\rho =\sum_{k}p_{k}\tau _{k}\otimes \mu _{k}\Rightarrow W_{\rho 
}\left( 
\mathbf{u}\right) =\sum_{k}p_{k}W_{\tau _{k}}\left( u^{(0)}\right) 
W_{\mu
_{k}}\left( u^{(1)}\right) .  \label{sepeqn}
\end{equation}

A natural definition of the characteristic function $\chi =\chi 
_{\rho }$ is
to use Eq.~(\ref{Smatrixpp}) with $n=2,$ and define 
\begin{equation} 
\widetilde{\chi }(\mathbf{w})=tr\left[ \rho S_{\mathbf{g}_{0}(\alpha
)}^{b_{0}}S_{\mathbf{g}_{1}(\alpha )}^{b_{1}}\right] 
\label{chitilde}
\end{equation}е
where $\mathbf{w}=b_{0}\mathbf{g}_{0}(\alpha 
)+b_{1}\mathbf{g}_{1}(\alpha ).$
We can rewrite the product of the $S$ matrices on $H_{p^{2}}$ as a 
direct
product of $S$ matrices on $H_{p}^{(0)}\otimes H_{p}^{(1)}.$  
Rewriting 
$\mathbf{w}=u^{(0)}\oplus u^{(1)}$ where $u^{(j)}\in V_{2}^{(j)}(p),$ 
we find 
\[
\widetilde{\chi} (\mathbf{w})=\eta ^{\Phi }tr\left[ \rho 
S_{u^{(0)}}\otimes
S_{u^{(1)}}\right] .
\]
The problem with this definition is that in general $\Phi \neq \phi 
(u^{(0)})+\phi
(u^{(1)}),$ so that the corresponding  Wigner function would not 
factor when $\rho $ is a
product state. Now as pointed out before, there is some freedom in the
choice of phase in defining the characteristic function and the Wigner
function. For this reason it is convenient to introduce a phase 
factor into
the definition of the characteristic function to avoid this problem. 
We
shall therefore define the characteristic function as 
\begin{equation}
\chi (\mathbf{w}) =\eta ^{-\Theta }\widetilde{\chi }(\mathbf{w)}%
,  \label{chip2}
\end{equation}
using Eq.~(\ref{chitilde}) and the $\Theta $
defined in Eq.~(\ref{phase}) that is linear in $b_{0}$ and  $b_{1}$. 
The linearity in the $b's$ is important, as we shall see, because we 
want to write the 
analog of Eq.~(\ref{discreteWigner1}) with the appropriate projection 
operators given in Eq.~(\ref{Projector2}).

The underlying reason for having to introduce the phases arises from 
the 
fact that we are using the
geometries of $V_{2}(p^{2})$ and $V_{4}(p)$. That fact forces us to go
into some detail to define appropriate phase factors and to confirm 
that
they work.

For example, consider the case of $p$ odd discussed at the end of the 
last
section. For $\alpha \neq p^{2}$ define 
\begin{equation}
\chi _{\rho }\left( \mathbf{w}\right) =Tr\left[ \rho \left( \eta
^{-Da_{1}}S_{\mathbf{g}_{0}}\right) ^{b_{0}}\left( \eta 
^{-Da_{1}}S_{\mathbf{%
g}_{1}}\right) ^{b_{1}}\right] ,  \label{charfunc2}
\end{equation}
\newline
and for $\alpha =p^{2}$ define 
\begin{equation}
\chi _{\rho }\left( \mathbf{w}\right) =Tr\left[ \rho \left( \eta
^{-2^{-1}}S_{\mathbf{g}_{0}}\right) ^{b_{0}}\left( \eta 
^{-2^{-1}}S_{\mathbf{g}%
_{1}}\right) ^{b_{1}}\right] .  \label{charfunc2a}
\end{equation}
Then define
\begin{equation}
W_{\rho }\left( \mathbf{u}\right) 
=\frac{1}{p^{4}}\sum_{\mathbf{w}}\eta ^{%
\mathbf{u}\circ \mathbf{w}}\chi _{\rho }\left( \mathbf{w}\right) .
\label{wigfunc2}
\end{equation}
Note again that $-2^{-1}$ is computed modulo $p$ and equals $\left(
p-1\right) /2.$ The vector symplectic product in the exponent of 
$\eta $ is
defined in Eq.~(\ref{syprmap}). From Eq.~(\ref{discreteWigner}) we 
can write
out the right hand side of Eq.~(\ref{wigsep2}) with one modification. 
For $%
mu_{p}=m\left( 0,1\right) $ take 
\begin{equation}
\chi _{\tau }( mu_{p}) =tr\left[ \tau \left( \eta
^{-2^{-1}}S_{u_{p}}\right) ^{m}\right] ,  \label{charfunc1A}
\end{equation}
as we did in Example 4 of Section \ref{SectionWigner}. Then $W_{\tau 
}\left(
u\right) $ is defined as the usual symplectic tranform and can be 
written as 
\begin{equation}
W_{\tau }\left( u\right) =\frac{1}{p^{2}}\left[ \sum_{m=0}^{p-1}\eta
^{u\circ mu_{p}}\chi _{\tau }\left( mu_{p}\right)
+\sum_{c=0}^{p-1}\sum_{m=1}^{p-1}\eta ^{u\circ mu_{c}}\chi _{\tau 
}\left(
mu_{c}\right) \right] .  \label{wigfunc1}
\end{equation}
Finally, we get the right hand side  of Eq.~(\ref{wigsep2}) as the 
trace of $\frac{1}{p^{2}}\left(
\tau \otimes \mu\right) $ times the expression 
\begin{eqnarray*}
rhs &=&\sum_{m_{0}=0}^{p-1}\sum_{m_{1}=0}^{p-1}\eta ^{\mathbf{u}\circ 
\left(
0,m_{0},0,m_{1}\right) }\eta ^{-2^{-1}\left( m_{0}+m_{1}\right)
}S_{m_{0}u_{p}}\otimes S_{m_{1}еu_{p}}+ \\
&&\sum_{\left( c_{0},c_{1}\right) \neq \left( p,p\right) }\sum_{\left(
m_{0},m_{1}\right) \neq \left( 0,0\right) }\eta ^{\mathbf{u}\circ 
\left(
m_{0}u_{c_{0}}\oplus m_{1}u_{c_{1}}\right) }\eta ^{-2^{-1}\left( 
\delta
\left( c_{0},p\right) m_{0}+\delta \left( c_{1},p\right) m_{1}\right)
}\left( S_{u_{c_{0}}}\right) ^{m_{0}}\otimes \left( 
S_{u_{c_{1}}}\right)
^{m_{1}}.
\end{eqnarray*}
The left hand side of Eq.~(\ref{wigsep2}) can be written as the trace 
of $%
\frac{1}{p^{2}}\left( \tau\otimes \mu\right) $ times the
expression
\begin{eqnarray*}
lhs &=&\sum_{b_{0}=0}^{p-1}\sum_{b_{1}=0}^{p-1}\eta ^{\mathbf{u}\circ 
\left(
b_{0}\mathbf{g}_{0}\left( p^{2}\right) +b_{1}\mathbf{g}_{1}\left(
p^{2}\right) \right) }\eta ^{-2^{-1}\left( b_{0}+b_{1}\right) 
}S_{\left(
0,b_{0},0,b_{1}\right) }+ \\
&&\sum_{\alpha \neq p^{2}}\sum_{\left( b_{0},b_{1}\right) \neq \left(
0,0\right) }\eta ^{\mathbf{u}\circ \left( b_{0}\mathbf{g}_{0}\left( 
\alpha
\right) +b_{1}\mathbf{g}_{1}\left( \alpha \right) \right) }\eta
^{-Da_{1}\left( b_{0}+b_{1}\right) }\left( S_{\mathbf{g}_{0}(\alpha
)}\right) ^{b_{0}}\left( S_{\mathbf{g}_{1}(\alpha )}\right) ^{b_{1}}.
\end{eqnarray*}
Note that in this equation we have the ordinary matrix product in the 
second
term. 

Our goal is to confirm that Eq.~(\ref{wigsep2}) holds with the above 
definitions of the characteristic functions. Using Eq.~(\ref{SPower}) 
we can pair the indices of the spin
matrices in Eq.~(\ref{wigsep2}) to obtain the \textit{index
equation} relating terms in \textit{rhs} to \textit{lhs},
\begin{equation}
\mathbf{w}=m_{0}u_{c_{0}}\oplus 
m_{1}u_{c_{1}}=b_{0}\mathbf{g}_{0}\left(
\alpha \right) +b_{1}\mathbf{g}_{1}\left( \alpha \right) ,  
\label{indexeqn}
\end{equation}
which includes the $\mathbf{w}=\left( 0,0,0,0\right) $ term that is
incorporated in the first summations. It follows that the phase 
factor $\eta
^{\mathbf{u}\circ \left( b_{0}\mathbf{g}_{0}\left( \alpha \right) 
+b_{1}%
\mathbf{g}_{1}\left( \alpha \right) \right) }$ is common to the
corresponding terms of $rhs$ and $lhs$, and we can cancel it. It is 
also
obvious that the $\alpha =p^{2}$ terms equal the corresponding terms
associated with $c_{0}=c_{1}=p$ and that the remaining phase factors 
in this
case are also equal if we set $m_{k}=b_{k}$.

To match terms in the second sets of summations, we multiply out the 
powers
of the spin matrices in \textit{lhs} to obtain 
\[
S_{m_{0}u_{c_{0}}}\otimes S_{m_{1}u_{c_{1}}}=S_{b_{0}\left( 
1,2a_{0}\right)
+b_{1}\left( 0,2Da_{1}\right) }\otimes S_{b_{0}\left( 0,2Da_{1}\right)
+b_{1}\left( 1,2Da_{0}\right) },
\]
where the equality follows from the index equation. This process 
introduces
phase factors using Eqs.(\ref{product}) and (\ref{SPower}), and it 
remains
to prove that the resulting exponents of $\eta $ are equal. 
Specifically,
one has to verify that subject to Eq.~(\ref{indexeqn}) 
\begin{equation}
\left( 1-\delta \left( c_{0},p\right) \right) c_{0}
\binom{m_{0}}{2}
 +\left( 1-\delta \left( c_{1},p\right) \right) c_{1}
\binom{m_{1}}{2}-2^{-1}\left( \delta \left( c_{0},p\right) 
m_{0}+\delta \left(
c_{1},p\right) m_{1}\right)   \label{leftphase1}
\end{equation}
equals 
\begin{equation}
-Da_{1}\left( b_{0}+b_{1}\right) +2a_{0}
\binom{b_{0}}{2}+2Da_{0} 
\binom{b_{1}}{2} +2b_{0}b_{1}Da_{1}.  \label{rightphase1}
\end{equation}
We verify the equality for $\alpha \neq p^{2}$ by considering 
different cases. Let $\alpha=a_{0}+a_{1}\lambda$. If $b_{0}$ and 
$b_{1}$ are both
non-zero, $m_{0}=b_{0},$ $m_{1}=b_{1}$ and 
\begin{eqnarray*}
b_{0}c_{0} &=&b_{0}2a_{0}+b_{1}2Da_{1}, \\
b_{1}c_{1} &=&\left( b_{0}2Da_{1}+b_{1}2Da_{0}\right) .
\end{eqnarray*}
If $b_{0}=0,$%
\[
\mathbf{w}=(0,b_{1}2Da_{1})\oplus (b_{1},b_{1}2Da_{0})
\]
and $c_{0}=p,$ $m_{0}=b_{1}2Da_{1},$ $c_{1}=2Da_{0}$ and 
$m_{1}=b_{1}.$
Similarly, for $b_{1}=0$ 
\[
\mathbf{w}=(b_{0},b_{0}2a_{0})\oplus (0,b_{0}2Da_{1}),
\]
and $m_{0}=2Da_{0},$ $m_{0}=b_{0},$ $m_{0}=p$ and 
$m_{1}=b_{0}2Da_{1}.$ Substituting these expressions in 
Eq.~(\ref{leftphase1}) gives (\ref{rightphase1}). We
have gone through this in some detail because the method illustrated
generalizes to the case of complete separability of $n$ subsystems. 
It 
should be noted that the argument leading to Eq.~(\ref{wigsep2}) did 
not 
require that $\tau$ or $\mu$ be a density matrix.

Our ability to add a phase factor to the definition of the 
characteristic
function is related to an arbitrariness in the assigning of state 
vectors in
a basis on the Hilbert space to lines in phase space as noted in 
\cite{Woottersphase}. This is 
illustrated in \ref{Ex2} below.

A different definition of the Wigner function in terms of the 
characteristic
function can be found in \cite{Vourdas}. Vourdas replaces the $M$
transformation by introducing the trace operation into the Fourier
transformation.

\subsection{\textbf{\ Properties of the Wigner Function}}

Because we have used the same format in defining the Wigner function 
for two
subsystems, Eq.~(\ref{wigfunc2}), as was used in defining it for a 
single
subsystem, Eq.~(\ref{discreteWigner}), we expect the properties in 
Section \ref
{WignerCont} to hold. With the definition of 
$\chi_{\rho}(\mathbf{w)}$ in  Eqs.~(%
\ref{charfunc2}) and (\ref{charfunc2a}) conditions (\ref{chi0}) and 
(\ref
{chiCC}) are satisfied. The discrete Wigner function $W_{\rho }$ for a
density $\rho $ on $H_{p^{2}}$ is defined using the symplectic Fourier
transform Eq.~(\ref{wigfunc2}); consequently, $W_{\rho }$ is convex 
linear on
the space of densities and linear on the space of $p^{2}\times p^{2}$
matrices. Again, the defining Eq.~(\ref{wigfunc2}) is invertible, so 
that one
can obtain the $\chi _{\rho }\left( \mathbf{w}\right) $ and thus the 
spin
coefficients of $\rho $ from the Wigner function. 
With this definition Plancheral's
formula becomes \[
p^{4}\sum_{\mathbf{v}\in V_{4}(p)}|W(\mathbf{v})|^{2}=|\chi
(0)|^{2}+\sum_{q,r=0}^{p}\sum_{M}|\chi (m^{(0)}u_{q}^{(0)}\oplus
m^{(1)}u_{r}^{(1)})|^{2}.
\]
We also have, as in Eq.~(\ref{quadratic}), that 
\[
tr\left[ \rho _{1}\rho _{2}\right] 
=p^{2}\sum_{\mathbf{v}}W_{\rho_{1}е}(\mathbf{v}%
)W_{\rho_{2}е}(\mathbf{v}),
\]
and, consequently, $|W(\mathbf{v})| \leq 1/p $ and  
$|suppW(\mathbf{v})|\geq 1/p^{2}$. Using the notation of Eq.~(%
\ref{Projector2}), we can write 
\begin{equation}
W_{\rho }\left( \mathbf{u}\right) =tr\left[ \rho A\left( 
\mathbf{u}\right)
\right]   \label{fouriercoef}
\end{equation}
where 
\begin{eqnarray}
p^{2}A\left( \mathbf{u}\right)  &=&-S_{0,0}\otimes 
S_{0,0}+P_{p^{2}}\left(
-2^{-1}+\mathbf{u}\circ \mathbf{g}_{0}\left( p^{2}\right) 
,-2^{-1}+\mathbf{u}%
\circ \mathbf{g}_{1}\left( p^{2}\right) \right)   \label{Adef2} \\
&&+\sum_{\alpha \neq p^{2}}P_{\alpha }\left( -Da_{1}+\mathbf{u}\circ 
\mathbf{%
g}_{0}\left( \alpha \right) ,-Da_{1}+\mathbf{u}\circ 
\mathbf{g}_{1}\left(
\alpha \right) \right) ,  \nonumber
\end{eqnarray}
corresponding to Eq.~(\ref{A operator}). From Eq.~(\ref{Adef2}) it 
follows
that $W_{\rho }$ is real for densities $\rho $. In particular, 
$\left\{ A(%
\mathbf{u})\right\} $ again defines a complete orthogonal set of 
Hermitian
matrices. The argument is analogous to that leading to Eq.~(27) and 
leads to 
\[
Tr\left[ A\left( \mathbf{u}\right) A\left( \mathbf{v}\right) \right]
=p^{-2}\delta \left( \mathbf{u},\mathbf{v}\right) 
\]
Thus we can interpret the Wigner function $W_{\rho}$ as the set of 
coefficients of $%
\rho $ in the orthogonal expansion relative to $\left\{ A(\mathbf{u}%
)\right\}$ analogous to Eq.~(\ref{inversion1}) .

The analogues of the other properties of Section \ref{Properties} 
follow in
the same way as before. $W$ is normalized since we can use Eq.~(\ref
{fouriercoef}) to prove 
\[
\sum_{\mathbf{u}}W_{\rho }\left( \mathbf{u}\right) =Tr\left[ \rho 
\sum_{%
\mathbf{u}}A\left( \mathbf{u}\right) \right] =Tr\left[ \rho \right] =1
\]
since Eq.~(\ref{Ieqn}) holds in this case.
If $\rho ^{\prime }=S_{\mathbf{z}}^{\dagger }\rho S_{\mathbf{z}}\in
H_{p}\otimes H_{p},$ then $\chi _{\rho ^{\prime }}\left( 
\mathbf{w}\right)
=\eta ^{\mathbf{z}\circ \mathbf{w}}\chi _{\rho }\left( 
\mathbf{w}\right) $,
and $W_{\rho ^{^{\prime }}}\left( \mathbf{u}\right) =W_{\rho }\left( 
\mathbf{%
u}+\mathbf{z}\right) $ as before.

Summing $W_{\rho }\left( \mathbf{u}\right) $ over a ``line'' in 
$V_{2}\left(
p^{2}\right) $ corresponds to summing over a translation of a two 
dimensional
subspace in $V_{4}\left( p\right) $ and again leads to a marginal
probability $Tr\left[ \rho P_{\alpha }\left( s_{1},s_{2}\right) 
\right] $.
To see this let $C_{\alpha }$ denote the two dimensional subspace 
associated
with $\alpha $. It is easy to show that 
\[
\sum_{\mathbf{u}\in C_{\alpha }}W_{\rho }\left( 
\mathbf{u}+\mathbf{r}\right)
=\frac{1}{p^{2}}\sum_{\mathbf{w}\in C_{\alpha }}\eta 
^{\mathbf{r}\circ 
\mathbf{w}}\chi _{\rho }\left( \mathbf{w}\right) . 
\]
This can be written as the trace of $\rho $ against the projection $%
P_{\alpha }\left( s_{1},s_{2}\right) $ for appropriate indices 
$s_{1}$ and $%
s_{2}$ which depend on $\mathbf{r}$ and the phase factors used in the
definition of the characteristic functions. Thus using definition of
Eq.~(\ref{wigfunc2}) the
Wigner function satisfies the conditions proved in section 
\ref{Properties}
and the requirement that $W_{\rho}$ factor for separable $\rho$ as in 
Eq.~(\ref{sepeqn}).

As pointed out to us by Wootters, Eq.~(\ref{wigsep2}) may be used to give a 
positive answer to a question posed in \cite{Woottersphase}. That is, 
with the phase factors given above, we have 
\[ 
A(\mathbf{u})=A(u^{(0)})\otimes A(u^{(1)}).
\]
where $\mathbf{u}=u^{(0)}\oplus u^{(1)}.$
The proof is easy, rewrite Eq.~(\ref{wigsep2}) as 
\[
tr\left[\tau_{0} \otimes \tau_{1}еA(\mathbf{u})\right]=
tr\left[\tau_{0} \otimes \tau_{1}еA(u^{(0)}\oplus A(u^{(1)})\right].
\]
This equality holds even the $\tau$'s are not densities. Since 
Hermitian matrices of the form $\tau \otimes \mu$ form a basis of 
$M_{p^{2}е}$, this inequality holds for all $A(\mathbf{u})$ for 
all $\mathbf{u}\in V_{4}(p)$.

\subsection{Examples}

\subsubsection{Maximally entangled state}

For prime $p$ let $\left| \Psi \right\rangle =\frac{1}{\sqrt{p}}
\sum_{j}\left| j\right\rangle \left| j\right\rangle $, so that $\rho 
\equiv
\left| \Psi \right\rangle \left\langle \Psi \right| 
=\frac{1}{p}\sum_{j,k}$ $%
\left| j\right\rangle \left\langle k\right| \otimes \left| 
j\right\rangle
\left\langle k\right| $. By the separability property and linearity 
we know
that if $\mathbf{u}=u^{(0)}\oplus u^{(1)}=\left(
x_{0},y_{0},x_{1},y_{1}\right) $, then 
\[
W_{\rho }\left( \mathbf{u}\right) =\frac{1}{p}\sum_{j,k}W_{\left|
j\right\rangle \left\langle k\right| }\left( u^{(0)}\right) W_{\left|
j\right\rangle \left\langle k\right| }\left( u^{(1)}\right) ,
\]
where $W_{\left| j\right\rangle \left\langle k\right|
}\left( u\right) $ is defined in Eq.~(\ref{pointwig}). It follows 
that  
\[
W_{\rho }\left( \mathbf{u}\right) =\frac{1}{p^{3}}\sum_{j,k}\eta 
^{\left(
x_{0}+x_{1}+1\right) \left( k-j\right) }\delta \left(
y_{0}+2^{-1}(j+k),0\right) \delta \left( y_{0}+2^{-1}\left( j+k\right)
,0\right) ,
\]
and simplifying we get 
\[
W_{\rho }\left( \mathbf{u}\right) =\frac{1}{p^{2}}\delta \left(
1+x_{0}+x_{1},0\right) \delta \left( y_{0},y_{1}\right) 
\]
Thus the Wigner function for this maximally entangled state equals 
$1/p^{2}$
for the $p^{2}$ four-vectors with $u^{(0)}=(x_{0},y_{0})$ and 
$u^{(1)}=
\left( -1-x_{0},y_{0}\right)$ and equals zero elsewhere.

Although the Wigner function for this state is positive, it is a 
non-classical state. In particular, entangled states violate Bell 
inequalities.  Since the Wigner function discussed in this example is 
not separable, it need not respect mathematical inequalities based on 
separability.

\subsubsection{\label{Ex2} MUB}

Let $\rho =P_{\alpha }\left(s_{0},s_{1}\right).$ In this
case it is simplest to use Eq.~(\ref{Adef2}) so that 
\begin{eqnarray*}
W_{\rho}( \mathbf{u})  &=&\frac{1}{p^{2}}tr\left[ P_{\alpha }\left(
s_{0},s_{1}\right) A(\mathbf{u})\right]  \\
&=&\frac{1}{p^{2}}\left[ -1+\sum_{\beta \neq \alpha }Tr\left[ 
P_{\alpha
}\left( s_{0},s_{1}\right) P_{\beta }\left( r_{0,\beta }\left( 
\mathbf{u}%
\right) ,r_{1,\beta }\left( \mathbf{u}\right) \right) \right] 
+Tr\left[
P_{\alpha }\left( s_{0},s_{1}\right) P_{\alpha }\left( r_{0,\alpha 
}\left( 
\mathbf{u}\right) ,r_{1,\alpha }\left( \mathbf{u}\right) \right) 
\right]
\right]  \\
&=&\frac{1}{p^{2}}\left[ -1+p^{2}/p^{2}+\delta \left( 
s_{0},r_{0,\alpha
}\left( \mathbf{u}\right) \right) \delta \left( s_{1},r_{1,\alpha 
}\left( 
\mathbf{u}\right) \right) \right]  \\
&=&\frac{1}{p^{2}}\delta \left( s_{0},r_{0,\alpha }\left( 
\mathbf{u}\right)
\right) \delta \left( s_{1},r_{1,\alpha }\left( \mathbf{u}\right) 
\right) 
\end{eqnarray*}

Thus, $W_{\rho }\left( \mathbf{u}\right) $ equals $1/p^{2}$ on those 
$p^{2}$
four-vectors which match the given phases and equals zero elsewhere. 
For $%
\alpha \neq p^{2},$ $r_{k,\alpha }\left( \mathbf{u}\right) 
=-Da_{1}+\mathbf{u%
}\circ \mathbf{g}_{k}\left( \alpha \right) ,$ and it can be shown 
easily
that the set of four-vectors satisfying those conditions is 
\[\left\{ \mathbf{u}=b_{0}\mathbf{g}_{0}\left( \alpha \right) +b_{1}%
\mathbf{g}_{1}\left( \alpha \right) +\left(
0,s_{0}+Da_{1},0,s_{1}+Da_{1}\right) :b_{0},b_{1}\in GF(p)\right\} . 
\]
That is, $W_{\rho }\left( \mathbf{u}\right) $ is constant on a shift 
of the
two-dimensional subspace indexed by $\alpha .$ An analogous result 
holds if $%
\alpha =p^{2}$, and, as expected, this parallels the situation when 
$n=1.$

\subsection{\label{p=2}Separability of the Wigner function for p=2}

When $p=2$ Eq.~(\ref{wigfunc2}) can be used to define the Wigner 
function with the definition of the characteristic function 
given in Eq.~(\ref{chi2sq}) below. Properties other than 
separability follow as before, but the analysis leading to 
separability for
$p$ odd does not work in this case. The discussion above 
made use of the existence of a quadratic non-residue $D$; however, 
for $p=2$ no 
such quantity exists. In addition 
we must include the factors of $\alpha_{2}=\alpha_{2}е(1,1)=-i$ 
defined at the end of 
\ref{GenerSpinMat}. 

Explicit forms of generating vectors are
\[
G_{a_{0},a_{1}}=\{(1,a_{1},0,a_{0}+a_{1}),(0,a_{0}+a_{1},1,a_{0})\}
\]
for $\alpha=a_{0}+a_{1}\lambda \in GF(2^{2})$, and 
\[
G_{4}=\{(1,0,0,0),(0,0,1,0)\}.
\]
For the case $\alpha\neq 2^{2}$, the analog of
Eq.~(\ref{charfunc2}) is
\begin{equation}    
\chi_{\rho}(\mathbf{w})=Tr[\rho(\alpha_{2}^{a_{1}}\eta^{r_{0}}S_{\mathbf{g}_{0}})^{b_{0}}
 (\alpha_{2}^{a_{0}}\eta^{r_{1}}S_{\mathbf{g}_{1}})^{b_{1}}]
    \label{chi2sq}
\end{equation}
where $r_{0}$ and $r_{1}$ depend on $a_{0}$ and $a_{1}$. It is 
convenient to write the index equation Eq.~(\ref{indexeqn}) in the 
form 
\[
\mathbf{w}=(b_{0},q_{0})\oplus(b_{1},q_{1})=b_{0}\mathbf{g}_{0}(\alpha)+
b_{1}\mathbf{g}_{1}(\alpha),
\]
then it is not difficult to show that for $(b_{0},b_{1})\neq(0,0)$
\begin{eqnarray*}
    a_{0} & = & b_{0}q_{0}+(b_{0}+b_{1})q_{1}  \\
    a_{1} & = & (b_{0}+b_{1})q_{0}+b_{1}q_{1}.
\end{eqnarray*}
This allows us to replace the sums in the Wigner function over 
$a_{0}$ and  $a_{1}$ by sums 
over $q_{0}$ and  $q_{1}$. Now we can write
\begin{equation}
     \chi_{\rho}(\mathbf{w})=\eta^{(b_{0}r_{0}+b_{1}r_{1})}
    \alpha_{2}^{(b_{0}q_{0}+b_{1}q_{1})}\eta^{b_{0}b_{1}(q_{0}+q_{1})}
    Tr[\rho S_{b_{0},q_{0}}\otimes 
    S_{b_{1},q_{1}}].
    \label{chi2sqa}
 \end{equation}
 As stated above, we require that the phase factors are linear in 
 the b's. In order to enforce this it is easy to show that if 
 $r_{0}=0$ and $r_{1}=a_{0}=b_{0}q_{0}+b_{1}(a_{0}+a_{1})$ the 
 exponent of $\eta$ is simply $b_{1}q_{1}$. This calculation makes 
 use of the binary arithmetic, in particular $b^{2}=b$.
    
  Finally, we find that for $\rho=\tau \otimes \mu $ the phase factor 
  $\eta^{b_{1}q_{1}}$ requires that we use different one particle 
  Wigner functions for the two particles. Equivalently,
    \[
 W_{\rho}(u^{(0)}\oplus u^{(1)})=W_{\tau}(u^{(0)})W_{\mu^{t}}(u^{(1)})
    \]
 where $\mu^{t}$ is the transpose of the qubit density matrix $\mu$. 
 If we had taken $r_{0}=a_{1}$ and $r_{1}=0$ the 
 transpose would have appeared on $\tau$, rather than on $\mu$.

\subsection{Separability and Partial Transposition}

A necessary condition for separability of a density matrix of a 
bipartite
system $\rho \in H_{p}\otimes H_{p}$ is the Peres condition 
\cite{Peres}.
That is, the density matrix must transform into a density matrix under
partial transpose 
\begin{equation}
PT:\text{ }\langle j_{0},j_{1}|\rho |k_{0},k_{1}\rangle \rightarrow 
\langle
j_{0},k_{1}|\rho |k_{0},j_{1}\rangle  \label{PT trans}
\end{equation}

The transpose of a spin matrix is given by $\left( S_{j,k}\right) 
^{t}=\eta
^{-jk}S_{j,p-k};$ consequently, under the $PT$ transformation 
\[
\chi (b_{0}u_{q}^{(0)}\oplus b_{1}u_{r}^{(1)})\rightarrow \eta
^{-rb_{1}}tr\left[ \rho S_{u_{q}^{(0)}}^{b_{0}}\otimes
S_{u_{p-r}^{(1)}}^{b_{1}}\right] .
\]
Therefore, 
\[
PT:\ W\left( \mathbf{u}\right) =W\left( u^{(0)},u^{(1)}\right)
\rightarrow W\left( u^{(0)},p-(u^{(1)}+1)\right) 
\]
Unfortunately, this is not very useful since proving that $W$ 
corresponds to
a density matrix is not simple, see \ref{positivity}.

\section{\label{SectWig3}Wigner function: $d=p^{n}$.}

The generalization to $p^{n}$ degrees of freedom, where $p$ is  
prime, is
based on the Galois field (see $GF(p^{n})$ \cite{PRMUB} and Appendix 
\ref{appMUB}).
Starting from Eqs.~(\ref{ueq}) and (\ref{Calpha}), the set of vectors 
in $C_{\alpha}$ defined on the phase space  
$V_{2}(p^{n})$ generates a MUB.
As before $\mathbf{u}$ denotes a vector in $V_{2n}\left(
p\right) =\bigoplus\limits_{j=0}^{n-1}V_{2}^{(j)}(p)$ that we also 
write as $\mathbf{u}=\bigoplus\limits_{j=0}^{n-1}u^{(j)}$ where 
$u^{(j)}\in V_{2}^{(j)}$. These indices define the tensor products of 
spin matrices by 
$S(\mathbf{u})=\otimes _{j=0}^{n-1}S_{u^{(j)}}$. We also use the 
vector
symplectic product introduced in Eq.~(\ref{syprmap}).  When $p=2$ we 
need the usual factor of $-i$ if $u^{(j)}=(1,1)$.

The basic structure of the classes of indices defined by the mapping 
$M$ 
is
discussed in section \ref{MutuallyUBII} and Appendix \ref{appMUB}.
Specifically, class $C_{\alpha }$ of $V_{2}\left( p^{n}\right) $ maps
onto an $n$-dimensional subspace of $V_{2n}\left( p\right) $.
Each subspace is spanned by a set of $n$ vectors
$G_{\alpha}$ as defined in Eq.~(\ref{Generators}) that depend
explicitly on the parameters $\alpha =($ $a_{0},a_{1}.\ldots 
,a_{n-1})$ in $%
GF\left( p\right)$ which define $\alpha $ in $GF\left( p^{n}\right)$ 
as a
vector over $GF(p).$ Since $\mathbf{u}\circ \mathbf{v}=0$ for any two 
vectors in $%
C_{\alpha }$, it follows $\mathbf{g}_{r}(\alpha) \circ 
\mathbf{g}_{s}(\alpha)=0$ for two generating vectors.

As in the case of $n=1$ and $n=2$, each non-zero vector in one of 
the $C_{\alpha}$ is mapped into a $\mathbf{w}\neq 
\mathbf{0} \in V_{2n}(p)$ that can be written uniquely as
\[
 \mathbf{w}=\sum_{j=0}^{n-1}b_{j}\mathbf{g}_{j}(\alpha). 
\]
Assume $p$ is odd. Following the paradigm established earlier, for a 
given density $\rho $
define 
\begin{equation}
\chi_{\rho}е(\mathbf{w}) =Tr\left[ \rho \left(\eta^{r_{0}} 
S_{\mathbf{g}_{0}(\alpha)}\right) ^{b_{0}}\cdots \left(\eta^{r_{n-1}}
S_{\mathbf{g}_{n-1}(\alpha)е}\right) ^{b_{n-1}}\right] . 
\label{FinalChi}
\end{equation}

A discrete Wigner function for a density $\rho $ on 
$H^{p^{n}}$ is defined  
\[
W_{\rho}е( \mathbf{u})=
\frac{1}{p^{2n}}\sum_{\mathbf{w}}\eta ^{\mathbf{u}\circ 
\mathbf{w}}\chi_{\rho}( \mathbf{w}) 
\]
where $\mathbf{u}\circ \mathbf{w}$ is defined in Eq.~(\ref{syprmap}).

It is not difficult to show that $W_{\rho}( \mathbf{u}) $ is real and 
$%
\sum_{\mathbf{u}}W_{\rho}( \mathbf{u}) =1.$ The proof is simply a 
matter 
of
keeping track of the various representations: 
\begin{eqnarray*}
W_{\rho}е( \mathbf{u}) &=&\frac{1}{p^{2n}}\left[ 1+\sum_{\alpha
}\sum_{(b_{0},\ldots ,b_{n-1})\neq(0,\ldots,0)}^{p-1}
\eta ^{\sum_{j} \mathbf{u}\circ(b_{j}\mathbf{g}_{j}(\alpha))}
Tr\left[ \rho (\eta^{r_{0}}S_{\mathbf{g}_{0}(\alpha)}) ^{b_{0}}\cdots 
(\eta^{r_{n-1}}S_{\mathbf{g}_{n-1}(\alpha)}) ^{b_{n-1}}\right] 
\right] \\
&=&\frac{1}{p^{n}}\left[ -1+\frac{1}{p^{n}}\sum_{\alpha }Tr\left[ \rho
\prod_{j}\sum_{b_{j}=0}^{{p-1}}\left( \eta ^{(\mathbf{u}\circ 
\mathbf{g}_{j}(\alpha)+r_{j})}
S_{\mathbf{g}_{j}(\alpha)}\right) ^{b_{j}}\right] \right] \\
&=&\frac{1}{p^{n}}\left[ Tr\left( \rho \left[ -I+\sum_{\alpha 
}P_{\alpha}
(\mathbf{u}\circ\mathbf{g}_{j}(\alpha)+r_{j})\right] \right) \right].
\end{eqnarray*}
 This immediately confirms that $W_{\rho}е$
is real and shows that $W_{\rho}е\left( \mathbf{u}\right) $ is the 
coefficient of the
Hermitian matrix $A_{\mathbf{u}}=\left(-I+\sum_{\alpha 
}P_{\alpha}(\mathbf{u}\circ\mathbf{g}_{j}(\alpha)+r_{j})\right)/p^{n}.$ 

For the normalization, summing over $\mathbf{u}$
is equivalent to summing over all of the vectors in
each $\alpha $ summand: \[
\sum_{\mathbf{u}}W_{\rho}е( \mathbf{u}) =\frac{1}{p^{n}}\left[
-p^{n}+\sum_{\alpha }Tr\left( \rho I\right) \right] =1 
\]
as required.  Again note that we inserted a factor of 
$\eta^{r_{k}}$ into the $g_{k}(\alpha)$
term to define a set of Wigner functions. This latitude of 
definition is exploited in the Appendix to give complete 
separabilty when $p$ is an odd prime. Furthermore, with this \textit{special} 
choice of phase factors, the analog of Eq.~(\ref{wigsep2}) holds and 
the generalization of the argument for $n=2$ gives 
\[
A(\mathbf{u})=\bigotimes_{j=0}^{n-1}A(u^{(j)}),
\]
where$\mathbf{u}=\bigotimes_{j=0}^{n-1}u^{(j)}.$е

For $p=2$ the same calculations apply provided factors of $-i$ are 
included where required.  However the methodology establishing 
separability fails 
for $n>2$, and as far as we can determine the Wigner function as 
defined above does not respect 
separability. 

\section{Dynamics}

For completeness, we conclude with a discussion of Hamiltonian 
dynamics in the present context. Starting from the Heisenberg - von
Neumann equation for a $d-$dimensional system.
\begin{equation}
\frac{d\rho }{dt}=i\left[ \rho ,H\right] =i\left( H\rho -\rho H\right) .
\label{one}
\end{equation}
($\hbar =1$) we obtain a closed form for the dynamics of
either the Wigner function or the characteristic function when $d=p,$ 
a prime.

Let $p$ denote an odd prime. The spin coefficients of a density $\rho $ are
defined by 
\begin{equation}
s_{u}=tr\left( S_{u}^{\dagger }\rho \right)  \label{two}
\end{equation}
so that 
\begin{equation}
\rho =\frac{1}{p}\left[ \sum_{u}s_{u}S_{u}\right] .  \label{three}
\end{equation}
In defining the Wigner function, however, we emphasized the role of the
characteristic functions $\chi _{\rho }\left( mu_{a}\right) $ rather than
the spin coefficients, and we also noted that one could add phase factors.
For this discussion we use 
\[
\chi _{\rho }\left( mu_{a}\right) =\left\{ 
\begin{array}{c}
\begin{array}{cc}
tr\left( \rho \left( \eta ^{2^{-1}a}S_{u_{a}}\right) ^{m}\right) & a\neq p
\end{array}
\\ 
\begin{array}{cc}
tr\left( \rho \left( S_{u_{p}}\right) ^{m}\right) & a=p
\end{array}
\end{array}
\right. 
\]
since the extra phase factors simplify the analysis. The same convention
will be used for the Hamiltonian $H$. Of course, the spin function and 
characteristic function are simply related.
Using Eq.~( \ref{discrete chi}), we obtain for $y=\left( y_{0},y_{1}\right) =mu_{a}$ 
\begin{equation}
s_{y}=\eta ^{2^{-1}y_{0}y_{1}}\chi _{\rho }\left( -y\right) ,  \label{four}
\end{equation}
and the phase factors allow us to avoid making $a=p$ an exceptional case in (%
\ref{four}). Thus (\ref{three}) becomes 
\begin{equation}
\rho =\frac{1}{p}\sum_{u}\eta ^{2^{-1}u_{1}u_{0}}\chi _{\rho }\left(
-u\right) S_{u}.  \label{five}
\end{equation}
Since the spin matrices are orthogonal, it is easy to show that
\begin{equation}
\frac{d\chi _{\rho }\left( -w\right) }{dt}=i\sum_{u}L\left( w,u\right) \chi
_{\rho }\left( -u\right)  \label{seven}
\end{equation}
where 
\begin{equation}
L\left( w,u\right) =\frac{1}{p}\chi _{H}\left( u-w\right) \left( \eta
^{2^{-1}w\circ u}-\eta ^{2^{-1}u\circ w}\right) .  \label{eight}
\end{equation}
Equation (\ref{four}) enables one to convert (\ref{eight}) to describe the
dynamics in terms of the spin coefficients rather than the characteristic
functions. It is easy to check that $L$ is a Hermitian operator indexed by $%
V_{2}\left( p\right) $, so that (\ref{seven}) can be solved in closed form.

The evolution of the system can also be expressed in terms of the evolution
of the Wigner functions. Using Eq.~(\ref{discreteWigner}) together with the results
above, we avoid explicit use of the $A\left( u\right) $ operators. Since the
Wigner function is real, using Eq.~(\ref{chiCC})) we can write
\begin{equation}
W_{\rho }\left( v\right) =\frac{1}{p^{2}}\sum_{w}\eta ^{v\circ w}\chi _{\rho
}\left( w\right) =\frac{1}{p^{2}}\sum_{w}\eta ^{w\circ v}\chi _{\rho }\left(
-w\right) .  \label{twelve}
\end{equation}
Then taking the time derivative, using (\ref{eight}) and then inverting (\ref
{twelve}) gives 
\begin{equation}
\frac{dW_{\rho }(v)}{dt}=i\sum_{v}\tilde{L}\left( v,y\right) W_{\rho }\left(
y\right),  \label{ten}
\end{equation}
where 
\begin{equation}
\tilde{L}\left( v,y\right) =\frac{1}{p}\left[ \eta ^{2v\circ y}\chi
_{H}\left( 2\left( y-v\right) \right) -\eta ^{2y\circ v}\chi _{H}\left(
2\left( v-y\right) \right) \right]  \label{nine}
\end{equation}
 is Hermitian on $V_{2}\left( p\right).$

This representation works best when the density $\rho $ evolves in the
convex hull of the MUB projections. As an example when $p=3$, let the
Hamiltonian be 
\[
H=\omega \left( S_{0,1}+S_{0,1}^{\dagger}е\right) 
\]
and take $\rho \left( 0\right) $ to be $P_{(1,1)}\left(
0\right) =\frac{1}{3}\left[ S_{0,0}+S_{1,1}+\eta S_{2,2}\right].$ Computing 
$L$ and finding its spectral decomposition leads to the expression of $\rho
\left( t\right) $ in terms of MUB projections as
\begin{eqnarray*}
     \rho(t) & = & \frac{1}{3}[ 
\left( 1+2\cos(\omega t) \right) P_{(1,1)}(0)+\left(1+2\cos (\omega t+2\pi/3)\right) 
P_{(1,0)}(1)\\
     &  & +\left(1+2\cos (\omega t+4\pi /3)\right) P_{(1,2)}(2)].
\end{eqnarray*}

In the special case of $p=2,$ the necessity of selectively introducing a
factor of $-i$ modifies the form of $L$. Any density $\rho $ can be written
as 
\begin{eqnarray*}
\rho &=&\frac{1}{2}\left[ \sigma _{0}+m_{x}\sigma _{x}+m_{z}\sigma
_{z}+m_{y}\sigma _{y}\right] \\
&=&\frac{1}{2}\left[
S_{0,0}+s_{0,1}S_{0,1}+s_{1,0}S_{1,0}+s_{1,1}S_{1,1}\right].
\end{eqnarray*}
where $s_{0,1}=$ $m_{x},$ $s_{1,0}=$ $m_{z}$ and $s_{1,1}=$ $-im_{y},$ and
the $m^{\prime }s$ are real with square sum less than or equal to $1.$
Defining the characteristic function as before, 
\[
\chi _{\rho }\left( j,k\right) =Tr\left[ \rho \left( \alpha
_{j,k}S_{j,k}\right) \right] , 
\]
we find $\chi _{\rho }\left( u\right) $ equals the corresponding $m$ and 
\[
s_{j,k}=\left( -i\right) ^{jk}\chi _{\rho }\left( j,k\right) . 
\]
Working through the differential equation leads to a similar form: 
\begin{equation}
\frac{d\chi _{\rho }\left( v\right) }{dt}=i\sum_{u}L\left( v,y\right)
\chi _{\rho }\left( y\right)  \label{eighteen}
\end{equation}
with $L$ a Hermitian matrix given by 
\begin{equation}
L\left( v,y\right) =\frac{1}{2}\chi _{H}\left( v+y\right) \left[ \left(
i\right) ^{y\circ v}-\left( i\right) ^{v\circ y}\right]. 
\label{nineteen}
\end{equation}
Thus the structure of $L$ is similar to the $p>2$ case but with powers of $i$
rather than powers of $\eta =-1$. That difference makes the corresponding
equation for the Wigner function more complicated, and we do not present it
here. Our conclusion is that the discrete Wigner function is not
particularly useful for studying the dynamics of a two-level system.

A similar approach works for $n$ systems, and we record the results for $%
n=2. $ For ${\bf u=}u^{\left( 0\right) }\oplus u^{\left( 1\right) }=b_{0}%
{\bf g}_{0}\left( \alpha \right) +b_{1}{\bf g}_{1}\left( \alpha \right) $
set 
\[
\chi _{\rho }\left( {\bf u}\right) =\left\{ 
\begin{array}{c}
\begin{array}{cc}
Tr\left[ \rho \left( \eta ^{2^{-1}y_{00}}S_{{\bf g}_{0}}\right) ^{b_{0}}\rho
\left( \eta ^{2^{-1}y_{11}}S_{{\bf g}_{1}}\right) ^{b_{1}}\right] & 
\alpha\neq
p^{2}
\end{array}
\\ 
\begin{array}{cc}
tr\left( \rho \left( S_{{\bf g}_{0}}\right) ^{b_{0}}\left( S_{{\bf g}%
_{1}}\right) ^{b_{1}}\right) & \alpha=p^{2}
\end{array}
\end{array}
\right. 
\]
where for $\alpha \neq p^{2}$ we use ${\bf g}_{0}\left( \alpha \right)
=\left( 1,y_{00},0,y_{01}\right) $ and ${\bf g}_{1}\left( \alpha \right)
=\left( 0,y_{10},1,y_{11}\right) .$ Recall that $y_{01}=y_{10}$. One can
then prove for all cases of $\alpha $ that 
\[
\chi _{\rho }\left( -{\bf u}\right) =\chi _{\rho }^{*}\left( {\bf u}\right) 
\]
and setting $u^{\left( k\right) }=\left( u_{0}^{\left( k\right)
},u_{1}^{\left( k\right) }\right) $ 
\begin{equation}
s_{{\bf u}}=\eta ^{2^{-1}\left( u_{0}^{\left( 0\right) }u_{1}^{\left(
0\right) }+u_{0}^{\left( 1\right) }u_{1}^{\left( 1\right) }\right) }\chi
_{\rho }\left( -{\bf u}\right) ,  \label{eleven}
\end{equation}
again for all $\alpha.$

Recall the vector symplectic product 
\[
{\bf u}\circ {\bf w=}\left( u^{\left( 0\right) },u^{\left( 1\right) }\right)
\circ \left( w^{\left( 0\right) },w^{\left( 1\right) }\right)
=\sum_{k=0}^{1}u^{\left( k\right) }\circ w^{\left( k\right) }, 
\]
and for convenience set $\left\langle {\bf u},{\bf u}\right\rangle
=u_{0}^{\left( 0\right) }u_{1}^{\left( 0\right) }+u_{0}^{\left( 1\right)
}u_{1}^{\left( 1\right) }.$ Then 
\begin{equation}
\rho =\frac{1}{p^{2}}\sum_{{\bf u}}\eta ^{2^{-1}\left\langle {\bf u},{\bf u}%
\right\rangle }\chi _{\rho }\left( -{\bf u}\right) S_{{\bf u.}}
\label{thirteen}
\end{equation}
Using the analogous representation for the Hamiltonian, we have the analogue of (\ref{seven}): 
\begin{equation}
\frac{d\chi _{\rho }\left( -{\bf w}\right) }{dt}=i\sum_{{\bf u}}L\left( {\bf %
w,u}\right) \chi _{\rho }\left( -{\bf u}\right) ,  \label{fourteen}
\end{equation}
where 
\begin{equation}
L\left( {\bf w,u}\right) =\frac{1}{p^{2}}\chi _{H}\left( -{\bf w+u}\right)
\left( \eta ^{2^{-1}{\bf w\circ u}}-\eta ^{2^{-1}{\bf u\circ w}}\right)
\label{fifteen}
\end{equation}
is Hermitian on $V_{4}\left( p\right) $.

The derivation of the dynamics in terms of the Wigner functions follows
almost word for word the pattern in the $n=1$ case, since the computation of
the Wigner function in terms of the characteristic function is symbolically
identical. This time

\begin{equation}
\tilde{L}\left( {\bf v},{\bf z}\right) =\frac{1}{p^{2}}\left[ \eta ^{2{\bf %
z\circ v}}\chi _{H}\left( 2{\bf v-2z}\right) -\eta ^{2{\bf v\circ z}}\chi
_{H}\left( 2{\bf z-2v}\right) \right]  \label{sixteen}
\end{equation}
is Hermitian on $V_{4}\left( p\right) $ and

\begin{equation}
\frac{dW_{{\bf v}}}{dt}=i\sum_{{\bf z}}\tilde{L}\left( {\bf v},{\bf z}%
\right) W_{{\bf z}}.  \label{seventeen}
\end{equation}

When $p=2=n,$ we obtain a structurally similar result, although as before
powers of $i$ appear instead of powers of $\eta $. Letting $u^{\left(
k\right) }=\left( u_{0}^{\left( k\right) },u_{1}^{\left( k\right) }\right) $
and ${\bf u}=u^{\left( 0\right) }\oplus u^{\left( 1\right) }$, 
\[
\chi _{\rho }\left( {\bf u}\right) =\left( -i\right) ^{\left\langle {\bf u},%
{\bf u}\right\rangle }s_{{\bf u}} 
\]
and 
\[
\frac{d\chi _{\rho }\left( {\bf w}\right) }{dt}=i\sum_{{\bf u}}\left( \frac{1%
}{4}\chi _{H}\left( {\bf w+u}\right) \left[ i^{u\circ w}-i^{w\circ u}\right]
\right) \chi _{\rho }\left( {\bf u}\right) . 
\]
The operator in the sum is Hermitian, and again the transformation to the
Wigner function context does not seem to be particularly useful.

\begin{acknowledgments}
 A preliminary version of this work was presented at the Feynman 
 Festival at the University of Maryland, College Park in August 2004.
This work was supported in part by NSF grants EIA-0113137 and 
DMS-0309042.
\end{acknowledgments}

\section{Appendices}

\subsection{\label{FiniteFields}Finite fields} Reference \cite{LN}

A finite field $K$ is a finite set of elements that contains an 
additive
unit $0$ and a multiplicative unit $1,$ that $K$  is an Abelian group 
with
respect to addition, $K^{*}=K-\{0\}$ forms an Abelian group under 
multiplication,
and the usual associative and distributive laws hold. The simplest 
example
of a finite field is the set of integers modulo a prime number $p$ 
that is
denoted by $Z_{p}=\{0,1,\cdots ,p-1\}.$  If $p$ is not prime there are
elements that do not have inverses, for example the set 
$Z_{4}^{*}=\{1,2,3\}$
does not form a multiplicative group because $2^{2}=0$ $\bmod{4}.$

It can  be shown that if $K$ is a finite field, then $\left| K\right| 
$, the
number of elements in $K$, is $p^{n}$, the power of a prime. Fields 
with the
same number of elements are isomorphic and are generically denoted as 
the
Galois field $GF\left( p^{n}\right) $. A field containing $p^{n}$ 
elements, 
$n>1,$ can be constructed using an irreducible polynomial $f$ of 
degree $n$
that has coefficients in $GF(p)=Z_{p}$. Let 
\[
f\left( x\right) =x^{n}+c_{n-1}x^{n-1}+\ldots +c_{1}x+c_{0} 
\]
be such a polynomial.  Let 
$\lambda \notin GF(p)$ denote a symbolic root of $f\left( x\right) 
=0$ so
that 
\begin{equation}
\lambda ^{n}=-\left( c_{n-1}\lambda ^{n-1}+\ldots +c_{1}\lambda
+c_{0}\right) .  \label{GFpoly}
\end{equation}
It can be shown that each element in $GF\left( p^{n}\right) $ can be 
represented
as 
\begin{equation}
\alpha \left( \lambda \right) =\sum_{k=0}^{n-1}a_{k}\lambda ^{k}.
\label{Polynomial}
\end{equation}
Addition and multiplication proceed in the usual manner with the 
replacement
of powers of $\lambda $ greater than $n-1$ reduced by using 
Eq.~(\ref{GFpoly}).
While the explicit representation depends on the choice of $f$, the 
theory
guarantees different representations are isomorphic.

As an example, we saw in Section \ref{MutuallyUBII} that if $n=p=2$, 
then $%
f\left( x\right) =x^{2}+x+1$ and $GF(4)=\left\{ 0,1,\lambda ,\lambda
+1\right\} $. For $p$ an odd prime and $n=2$ we noted that elements 
of $%
GF(p^{2})$ could be written as $j+k\lambda $, where $j$ and $k$ are 
in $%
GF(p) $ and $f(x)=x^{2}-D$ with $D$ a quadratic non-residue 
$\bmod{p}$.

In addition, there is a \textit{trace} operation defined on $GF\left(
p^{n}\right) $ that is linear over $GF\left( p\right) $ and that maps 
$
GF\left( p^{n}\right) $ to $GF\left( p\right) $. Specifically, if 
$\lambda
_{0},\ldots ,\lambda _{n-1}$ denote the $n$ distinct roots of $f$, 
then 
\[
tr\left( \alpha \left( \lambda \right) \right) \equiv 
\sum_{r=0}^{n-1}\alpha
\left( \lambda _{r}\right) . 
\]
 The elements $\alpha $ in $GF(p^{n})$ can thus be viewed as a vector 
space
over the field $GF(p)$ with basis $\left\{ \lambda ^{k}:0\leq 
k<n\right\} $.
A \textit{dual basis} $\left\{ g_{k}\left( \lambda \right) :0\leq 
k<n\right\} $ can
be defined such that elements of $GF(p^{n})$ also can be written as a 
linear
combinations of the $g_{k}$'s with coefficients in $GF(p)$. The 
definition
of a dual basis uses the trace operation with the requirement that
\[
tr\left[\lambda ^{j}g_{k}\left( \lambda \right) \right] =\delta \left(
j,k\right). 
\]
This structure was described in the Appendix of \cite{PRMUB} and the 
complete theory is presented in \cite{LN}.

\subsection{\label{appMUB}Mutually unbiased bases for d=p$^{n}.$}

For the finite field $GF(p^{n}),$ as is explained in section \ref
{MutuallyUBII}, we start with a vector space $V_{2}(p^{n}).$ 
We need to map the vectors in $V_{2}(p^{n})$ onto the space $%
V_{2n}(p)$ in order to write out the spin matrices corresponding to 
the set
of MUB. A typical
vector $\beta u_{\alpha }$ can be written as 
\begin{equation}
\beta u_{\alpha }=\sum_{j=0}^{n-1}\left( x^{(j)}(\alpha ,\beta
)e_{j}+y^{(j)}(\alpha ,\beta )f_{j}\right) .  \label{betau}
\end{equation}
The $x^{(j)}\left( \alpha ,\beta \right) $ and $y^{(j)}\left( \alpha 
,\beta
\right) $ are in $GF\left( p\right) $ and $\left\{ e_{j},\ f_{k}:0\leq
j,k<n\right\} $ is a set of $2n$ linearly independent vectors over 
$GF\left(
p^{n}\right) .$ It is convenient to take them to be of the form $%
e_{j}=\lambda ^{j}\left( 1,0\right) $ and $f_{k}=g_{k}(\lambda )\left(
0,1\right) $ so that
\[
tr\left( f_{k}\circ e_{j}\right) =tr\left( \lambda ^{j}g_{k}\left( 
\lambda
\right) \right) =\delta \left( j,k\right) \text{.} 
\]

The key point to defining a MUB is that for two non-zero vectors in $%
V_{2}\left( p^{n}\right) $, say $\gamma _{1}u_{\alpha }$ and $\gamma
_{2}u_{\beta },$ $\gamma _{1}u_{\alpha }\circ \gamma _{2}u_{\beta 
}=0$ iff $%
\alpha =\beta .$ Consequently, if in Eq.~(\ref{betau}) we set $%
x_{r}^{(j)}=x^{(j)}(\alpha ,\beta _{r})$ and 
$y_{r}^{(j)}=y^{(j)}(\alpha
,\beta _{r})$ for $r=1$ and $2$, we have 
\begin{eqnarray}
0 &=&tr\left( \beta _{1}u_{\alpha }\circ \beta _{2}u_{\alpha }\right) 
\nonumber \\
&=&tr\left( \sum_{j=0}^{n-1}\sum_{k=0}^{n-1}\left(
x_{1}^{(j)}e_{j}+y_{1}^{(j)}f_{j}\right) \circ \left(
x_{2}^{(k)}e_{k}+y_{2}^{(k)}f_{k}\right) \right)  \nonumber \\
&=&\sum_{j=0}^{n-1}\left(
y_{1}^{(j)}x_{2}^{(j)}-x_{1}^{(j)}y_{2}^{(j)}\right)  \nonumber \\
&=&\sum_{j=0}^{n-1}\left( \left( x_{1}^{(j)},y_{1}^{(j)}\right) \circ 
\left(
x_{2}^{(j)},y_{2}^{(j)}\right) \right) .  \label{sympprodmap}
\end{eqnarray}
Identifying the $j$th vector as the indices of the $j$th spin matrix 
in an $%
n $-fold tensor product, we have a necessary 
and                                                                                                                                                                                

sufficient condition for commutativity: 
\[
\otimes _{j=0}^{n-1}S_{x_{1}^{(j)},y_{1}^{(j)}е}\otimes
_{k=0}^{n-1}S_{x_{2}^{(k)}е,y_{2}^{(k)}е}=\otimes 
_{k=0}^{n-1}S_{x_{2}^{(k)}е,y_{2}^{(k)}е}\otimes
_{j=0}^{n-1}S_{x_{1}^{(j)}е,y_{1}^{(j)}е}. 
\]
Thus the set of $p^{n}$ vectors $\{\gamma u_{\alpha },\ \gamma \in
GF(p^{n})\}$ corresponds to a commuting class $\mathfrak{M}_{\alpha 
}$ of $p^{n}$
tensor products of spin matrices. 
The linear mapping $M:V_{2}\left( p^{n}\right) \rightarrow
V_{2n}\left( p\right) $ defined by 
\begin{equation}
M\left( \sum_{j}(x^{(j)}e_{j}+y^{(j)}f_{j})\right) 
=\left( x^{(0)},y^{(0)},\ldots,x^{(n-1)},y^{(n-1)}ееее \right)  
\label{M transform}
\end{equation}
is one-to-one and onto. Using Eq.~(\ref{sympprodmap})
this partitions the generalized spin matrices
into $d+1$ commuting classes having only the identity in common and 
satisfying the condition for the
existence of a set of $d+1$ mutually unbiased bases. In writing the 
$M$ mapping we are using a 
different definition of the basis $\{e_{j},f_{j}\}$
that the one used in \cite{PRMUB}. The definition in this paper
lends itself more readily to a discusion of separability.

\subsection{Separability and the M mapping}

We provide some details about the mapping $M:V_{2}\left( p^{n}\right)
\rightarrow V_{2n}\left( p\right) .$ Let $\lambda $ denote a root of 
an $nth$
order irreducible polynomial over $GF(p)$. On $V_{2}(p^{n})$ recall 
the set of vectors 
\[
\{e_{j}=\lambda ^{j}(1,0),\text{ }f_{j}=g_{j}(\lambda )(0,1),\text{ }
j=0,1,\cdots ,n-1\} 
\]
where $tr(f_{j}\circ e_{k})=\delta (j,k).$ Let $\alpha
=\sum\limits_{j=0}^{n-1}a_{j}\lambda ^{j}\in GF(p^{n})$ and using 
Eq.~(\ref
{betau}) define 
\begin{eqnarray}
u_{\alpha } &=&(1,\alpha 
)=e_{0}+\sum\limits_{j=0}^{n-1}y_{j}^{(0)}(\alpha
)f_{j}  \nonumber \\
y_{j}^{(0)}(\alpha ) &=&\sum\limits_{k=0}^{n-1}\left( tr\lambda
^{j+k}\right) a_{k.}  \label{app M map 1}
\end{eqnarray}
Then for $l=1,\cdots,p-1$%
\[
\lambda ^{l}u_{\alpha } =e_{l}+\sum_{j=0}^{n-1}\text{ 
}y_{j}^{(l)}(\alpha
)f_{j}  
\]
where
\begin{equation}
y_{j}^{(l)}(\alpha ) =\sum_{k=0}^{n-1} tr\left(\lambda ^{j+l+k}\right)
a_{k}.  \label{app M map 2}
\end{equation}

Let us work out the details for the case $p$ an odd prime and $n=2.$ 
We
choose as our irreducible polynomial $x^{2}-D=0$ $mod$ $p,$ where $D$ 
is a
quadratic non-residue. The symbolic roots of this equation are 
$\lambda $
and $(p-1)\lambda .$ For example if $p=3$ we may take $D=2.$ Then $%
tr[f(\lambda )]=f(\lambda )+f(2\lambda ).$ It is not difficult to 
show $%
g_{0}(\lambda )=2^{-1}$ and $g_{1}(\lambda )=(2D)^{-1}\lambda .$ Then 
\[
y_{0}^{(0)}=2a_{0},\text{ }y_{1}^{(0)}=y_{0}^{(1)}=2Da_{1},\text{ }
y_{1}^{(1)}=2Da_{0}. 
\]

We now can define the index generators of the MUB by 
\begin{eqnarray}
G_{\alpha } &=&\left\{ \mathbf{g}_{r}(\alpha )=M(\lambda ^{r}u_{\alpha
})=\bigoplus_{j=0}^{n-1}u_{r}^{(j)}(\alpha ),\quad u_{r}^{(j)}(\alpha
)=(\delta (j,r),y_{r}^{(j)}),\text{ }r=0,1,\cdots ,n-1\right\} 
\nonumber\\
G_{p^{n}} &=&\left\{ 
\mathbf{g}_{r}(p^{n})=\bigoplus_{j=0}^{n-1}u_{r}^{(j)}(p^{n}),\quad
u_{r}^{(j)}(p^{n})=(0,\delta (j,r)),\text{ }r=0,1,\cdots ,n-1\right\} 
\label{appGen} .
\end{eqnarray}
We should note that $\mathbf{g}_{r}(p^{n})$ is not 
$M(\lambda^{r}(0,1))$ but rather $M(g_{r}(\lambda)(0,1))$.ееее
For the example of odd $p$ and $n=2$ we find for $\alpha 
=a_{0}+a_{1}\lambda
,\, a_{0},a_{1}\in GF(p),$%
\begin{eqnarray}
G_{\alpha } &=&\left\{ \mathbf{g}_{0}(\alpha )=(1,2a_{0})\oplus 
(0,2Da_{1}),%
\text{ }\mathbf{g}_{1}(\alpha )=(0,2Da_{1})\oplus (1,2Da_{0})\right\} 
\nonumber \\
G_{p^{2}} &=&\left\{ \mathbf{g}_{0}(p^{2})=(0,1)\oplus 
(0,0),\mathbf{g}
_{1}(p^{2})=(0,0)\oplus (0,1)\right\}
\label{Genpsquared}
\end{eqnarray}
Each generator set is characterized by two independent four-vectors 
that
determine a plane containing $p^{2}$ points. These planes intersect 
at only
one point, the origin, and so the $p^{2}+1$ sets determine $p^{2}-1$ 
distinct
points and, including the origin, every point of $V_{4}(p).$

We note from Eqs.~(\ref{app M map 1}) and (\ref{app M map 2}) that $%
y_{k}^{(j)}=y_{j}^{(k)}$ which ensures the symplectic product is 
preserved
by the mapping. Therefore, we have for the general case
\begin{equation}
\lambda ^{r}u_{\alpha }\in V_{2}(p^{n})\rightarrow
\bigoplus_{j=0}^{n-1}u_{r}^{(j)}(\alpha )\in V_{2n}(p)\rightarrow 
S_{\mathbf{%
g}_{r}(\alpha )}^{b}\equiv 
\bigotimes\limits_{j=0}^{n-1}S_{u_{r}^{(j)}}^{b}
\label{app Gen S}
\end{equation}
where $u_{r}^{(j)}$ depends on $\alpha $ and the $b_{j}.$ With this
notation, the mapping from the index space to the spin matrices is 
complete, 
\[
G_{\alpha }\rightarrow \mathfrak{M}_{\alpha}=\left\{
\prod\limits_{r=0}^{n-1}S_{\mathbf{g}_{r}(\alpha
)}^{b_{r}}=\bigotimes\limits_{j=0}^{n-1}\prod%
\limits_{r=0}^{n-1}S_{u_{r}^{(j)}}^{b_{r}},\quad b_{r}\in 
GF(p)\right\} . 
\]
For the case of an odd prime $p$ and $n=2$ this result is  Eq.~(\ref
{SpinMatpsquar}). 

The spin matrices can be further expanded with the help of
Eqs.~(\ref{product}) and (\ref{SPower}), the symmetry of the 
$y_{r}^{(j)}(\alpha ),$ and a lot of algebra. First
\begin{eqnarray}
\bigotimes\limits_{j=0}^{n-1}\prod\limits_{r=0}^{n-1}S_{u_{r}^{(j)}(\alpha)}^{b_{r}}
&=&\bigotimes\limits_{j=0}^{n-1}S_{b_{j},q_{j}(\alpha 
)}\eta
^{\Phi _{j}(\alpha,b )}\label{Sprod}\\
q_{j}(\alpha,b)&=& \sum_{r=0}^{n-1}b_{r}y_{j}^{(r)}(\alpha ), 
\nonumber \\
\Phi _{j}(\alpha,b ) &=&\left(
2^{-1}b_{j}(b_{j}-1)y_{j}^{(j)}(\alpha
)+b_{j}\sum_{r=0}^{j-1}b_{r}y_{j}^{(r)}(\alpha )\right). 
\label{phisum}
\end{eqnarray}

If $b_{j}\neq 0$, define 
$q_{j}(\alpha,b)=b_{j}q_{j}^{\prime}(\alpha,b)$ and we have
\[
S_{b_{j},q_{j}(\alpha,b)}=S_{1,q_{j}^{\prime}(\alpha,b)}^{b_{j}}\eta^{-2^{-1}b_{j}(b_{j}-1)q_{j}^{\prime}(\alpha,b)}.
\]
If $b_{j}=0$, we have
\[
S_{0,q_{j}(\alpha,b)}=S_{0,1}^{q_{j}(\alpha,b)}
\]
After some manipulation, we can then rewrite Eq.~(\ref{Sprod}) as 
\begin{equation}   
\bigotimes\limits_{j=0}^{n-1}\prod\limits_{r=0}^{n-1}S_{u_{r}^{(j)}(\alpha         
)}^{b_{r}}=\eta^{\Theta(\alpha,b)} \bigotimes_{b_{j}\neq 0} 
S_{1,q_{j}^{\prime}(\alpha,b)}^{b_{j}}\bigotimes_{b_{j}= 
0}S_{0,1}^{q_{j}(\alpha,b)}
    \label{Sprod1}
\end{equation}
where the proper order of the tensor products is understood and where
\begin{equation}
    \Theta (\alpha ,b) =2^{-1}\sum_{r}b_{r}е\sum_{j\neq 
    r} y_{j}^{(r)}(\alpha)-2^{-1}\sum_{b_{j}=0}q_{j}(\alpha,b).
    \label{theta}
   \end{equation} 
   We now can incorporate the factor $\Theta$ into the definition 
   of $\chi$ as is done in Eqs.~(\ref{chip2}) and (\ref{FinalChi}).  
   Again leaving the ordering of the tensor products
   understood, Eq.~(\ref{Sprod1}) can be rewritten as
   \begin{equation}
    \prod_{r}\left( \eta^{-2^{-1}\sum_{j\neq r}y_{j}^{(r)}(\alpha)}
     S_{\mathbf{g}_{r}(\alpha)}\right)^{b_{r}}=\bigotimes_{b_{j}\neq 
0}     
S_{1,q_{r}^{\prime}(\alpha,b)}^{b_{j}}\bigotimes_{b_{j}=0}\left(\eta^{-1/2}     
S_{0,1}е\right)^{q_{j}}.  
    \end{equation}
 Therefore, we have shown that by introducing an appropriate phase 
    factor that depends on $r$ and $\alpha$ with each 
    $S_{\mathbf{g}_{r}(\alpha)}$ and by using $\eta^{-2^{-1}}S_{0,1}$ 
    in the definition of the one particle Wigner function, we can 
    define a Wigner function for all $n>1$ that respects complete 
    separabilty for odd $p$.
 Note that the spin
matrices appearing in the direct product are all in the standard form 
$S_{u_{c}}$
where $c\in I_{p}.$ 

For the example of odd prime $p$ and $n=2$ we have for $b_{0}$ and 
$b_{1}$ not equal to zero
\begin{eqnarray}
q_{0}(\alpha ,b) &=&b_{0}2a_{0}+b_{1}2Da_{1},\quad q_{1}(\alpha
,b)=b_{0}2Da_{1}+b_{1}2Da_{0}  \nonumber \\
\Theta (\alpha ,b) &=&(b_{0}+b_{1})2Da_{1}.  \label{phase}
\end{eqnarray}

As stated in Section \ref {p=2} the analysis for $p=2$ requires 
special handling.  For the case of a bipartite system, it was shown 
in 
\ref{p=2} that we could still prove a form of separability; however, 
for $n>2$ 
we have been unable to make the method used here work.

\subsection{Symplectic structure of the MUB}

We have seen that Eq.~(\ref{appGen}) determines the index sets for the
MUB. If $u,v$ $\in $ $V_{2}\left( p^{n}\right),$ consider the
transformations $A:$ $V_{2}(p^{n})\rightarrow V_{2}(p^{n})$ that 
leave the
symplectic product $u\circ v$ invariant. This is the set of $2\times 
2$
matrices with entries in $K=GF(p^{n})$ with unit determinant which 
forms the
symplectic group $Sp(2,K)$ \cite{Vourdas,Fivel,Hua}.

We now want to study the mapping $M$ defined in section 
\ref{MutuallyUBII}.
For simplicity we take $n=2$ so that the sets of generators of the 
MUB on $%
V_{4}(p)$ are $\{\mathbf{g}_{0}е(\alpha ),\mathbf{g}_{1}е(\alpha 
)\}.$ Introduce the $2\times 2$
matrix $\sigma =\left( 
\begin{array}{ll}
0 & -1 \\ 
1 & 0
\end{array}
\right) $ and the $4\times4$ matrix $J=\left( 
\begin{array}{ll}
\sigma & 0 \\ 
0 & \sigma
\end{array}
\right) $, then we can write the symplectic product in terms of an 
ordinary
inner product, $\mathbf{g}_{0}(\alpha )\circ \mathbf{g}_{1}(\alpha 
)=(\mathbf{g}_{0}(\alpha ),J\mathbf{g}_{1}е(\alpha )).$ Let 
$A\rightarrow A_{4}$ where $A_{4}$ is a linear transformation on 
$V_{4}(p)$, such that $A_{4}е\mathbf{g}_{0}е(\alpha
)\circ A_{4}е\mathbf{g}_{1}е(\alpha )=\mathbf{g}_{0}е(\alpha )\circ 
\mathbf{g}_{1}е(\alpha ).$ Then a matrix
representation of $A_{4}е$ must satisfy $A_{4}е^{t}JA_{4}е=J,$ where 
$A_{4}е^{t}$ is the
transpose of $A_{4}.$ The set of linear transformations that satisfy 
this
condition forms the symplectic group $Sp(4,Z_{p}).$ This is  
analogous to the canonical transformations for the continuous case. 
Under 
such a
transformation, the classes $C_{\alpha}$ determining the ONB of a 
given 
MUB are mapped into one another.  In summary, the symplectic group 
$Sp(2,p^{n})$ can be mapped onto a symplectic group $Sp(2n,Z_{p})$ 
and the operators $A_{2n}$ act on the bases in a MUB in such 
away as to leave the MUB invariant. For 
further discussion of the symplectic group in this context see 
\cite{Vourdas}.

\subsection{\label{Geometry}Phase Space and Finite Geometry.}

The purpose of this section is to review the role played by the 
geometry of the phase space. In Section \ref{MutuallyUB} we defined 
$V_{2}\left( p\right) $ to be the
phase space for the discrete Wigner function when $n=1$, and lines in
the vector space play an important role in relating the Wigner 
function to
probability measurements. 
By analogy, for $d=p^{n}$ a natural candidate for
phase space for a $d$ -level system is a two dimensional vector space 
with
entries from an appropriate set of scalars which has $d$ elements in 
it;
that is, we consider $V_{2}\left( p^{n}\right) =\left\{ \left( \alpha 
,\beta
\right) :\alpha ,\beta \in GF(p^{n)}е\right\}.$ However, in analogy 
with the continuous case for $n$ each described on a Hilbert space 
$H_{p}$ we use $V_{2n}(p)$ as the phase space. The $M$ mapping takes 
lines in $V_{2}(p^{n})$ to hyperplanes in $V_{2n}(p)$.

If $K$ denotes a finite field, the definition of a line in 
$V_{2}\left( K\right) $ is the obvious one$.$ A
line $L$ in $V_{2}\left( K\right) $ is a set of points in $V_{2}(K)$%
\[
\left\{ \left( x,y\right) :-\lambda y+\mu x+\gamma =0\text{ }x,y\in
K\right\} . 
\]
We always omit the case in which $\lambda =\mu  =0.$ It is important
to note the line consists of these points and only these points. For
example in $V_{2}(3),$ the sets $L_{1}=\{(0,0),$ $(1,1),$ $(2,2)\}$, 
$%
L_{2}=\{(0,0),(1,2),(2,1)\},$ and $L_{3}=$ $\left\{
(0,1),(1,2),(2,0)\right\} $ are lines. Two lines intersect only if 
they have
a point in common, otherwise they are parallel. The lines $L_{1}$ and 
$L_{2}$ in
the above example intersect at the origin, while $L_{1}$ and $L_{3}$ 
are
parallel.

$V_{2}\left( K\right)$ is also an example of an \textit{affine 
plane},  
a concept defined axiomatically in terms of a finite number of
points, a finite number of lines, and the relationship that a point 
lies on
a line. It can be shown that if a finite affine plane \textit{AP} 
exists,
then there is an $m$ such that \textit{AP} has exactly $m^{2}$ 
points, $%
m^{2}+m$ lines, each line contains $m$ points and each point is on 
$m+1$
lines. Two lines are said to be \textit{parallel} if they have no 
point in
common, and there are $m+1$ sets of $m$ parallel lines. (See 
\cite{LN} for a
summary of these results and references.) Since no affine plane is 
known for
an $m$ which is not a power of a prime, we are again restricted to 
dimension 
$p^{n}$.

The image under $M$ of lines in $V_{2}\left( p^{n}е\right),$ play a 
central
role in the definition of a Wigner function, and we summarize a few 
of their
properties. Using the generalization of Eq.~(\ref{line}) with 
$GF(p^{n})$ replacing $Z_{p}$ we have:
\begin{eqnarray*}
     L(\alpha,\gamma )&=&\left\{ xu_{\alpha }+\gamma u_{d}:x\in 
GF(p^{n})е\right\}  \\
    L(p^{n},\gamma )&=&\left\{ yu_{d}+\gamma u_{0}:y\in 
GF(p^{n})\right\}
\end{eqnarray*}
with $\alpha,\gamma\in GF(p^{n})$.  Recall that $u_{\alpha }=\left( 
1,\alpha
\right) ,$ and $u_{p^{n}е}=\left( 0,1\right).$  The vectors 
$u_{\alpha }$ and $u_{d}$ 
multiplying the variables $x$ or $y$ were
introduced earlier in Eq.~(\ref{ueq}) as a convenience. They now are
playing the role of ``slopes'' in an indexing of lines in $V_{2}\left(
p^{n}\right) $, a much more general setting. For each slope, as 
$\gamma $ 
varies over $GF(p^{n})$ we get
a set of parallel lines that contains each point in $V_{2}(p^{n})$ 
once.\\
1. Each line contains $p^{n}е$ elements, and there are $p^{2n}+p^{n}$ 
distinct 
lines.\\
2. The lines through the origin, $L(\alpha ,0)$ where $\alpha \in $ 
$I_{p^{n}}=GF(p^{n})\cup \left\{ p^{n}е\right\}$, only intersect at 
the origin.
Furthermore, 
\[
\bigcup_{\alpha \in I_{p^{n}е}}\left( L(\alpha ,0)-\left\{ \left( 
0,0\right)
\right\} \right) =V_{2}\left( p^{n}е\right) -\left\{ \left( 0,0\right)
\right\} . 
\]
3. Each set of parallel lines partitions $V_{2}(p^{n})$:
\[
V_{2}\left( p^{n}е\right) =\bigcup_{\gamma \in GF(p^{n})} L(\alpha 
,\gamma) 
\]
for each $\alpha \in I_{p^{n}е}.$

The relevance to this paper of the affine plane is that it can be 
shown that
for certain values of $d$, such as $d=6$, there is no corresponding 
affine
plane, and for other values of $d$, such as $d=12$, the existence of a
corresponding affine plane is an open question. (See standard texts in
combinatorics for more details or \cite{Woottersphase} for 
references.) 
We have already noted that if $K$ denotes a finite field, then 
$V_{2}\left(
K\right) $ is an example of an affine plane, so we are working in the 
most general
context with the necessary structure.

The last tool we need is the symplectic product of vectors
in $V_{2}\left( K\right) $ over the finite field $K$. Specifically,
recall that
\begin{equation}
\left( \mu _{1},\nu _{1}\right) \circ \left( \mu _{2},\nu _{2}\right) 
=\nu
_{1}\mu _{2}-\mu _{1}\nu _{2},  \label{symplectic1}
\end{equation}
where the algebra is in the field $K$. As an example, $u_{\alpha 
}\circ
u_{\beta }=0$ if and only if $\alpha =\beta $.
For each $\alpha \neq d $ in $I_{d}$, $\left( \lambda ,\mu \right) $ 
is on the line 
$L(\alpha ,\gamma )$ where $\gamma =$ $\left( \lambda ,\mu \right) 
\circ
u_{a}$.

Finally, from each $u_{\alpha}$ we generate $n$ linearly independent 
vectors that are mapped into an $n$-dimensional hyperplane in 
$V_{2n}(p)$ using Eqs.~(\ref{app M map 1}) and(\ref{app M map 2}).

\subsection{Examples of the geometry}

\subsubsection{One qubit}

Let $K=GF\left( 2\right) $, the Galois field consisting of the 
integers $%
\bmod \: 2$. The six lines of $V_{2}\left( 2\right) $ fall into three
classes containing two parallel lines: 
\begin{eqnarray*}
&&\left\{ L(0,0)=\left\{ \left( 0,0\right) ,\left( 1,0\right) \right\}
,L(0,1)=\left\{ \left( 0,1\right) ,\left( 1,1\right) \right\} 
\right\} \\
&&\left\{ L(1,0)=\left\{ \left( 0,0\right) ,\left( 1,1\right) 
\right\} \text{
, }L(1,1)=\left\{ \left( 0,1\right) ,\left( 1,0\right) \right\} 
\right\} \\
&&\left\{ L(2,0)=\left\{ \left( 0,0\right) ,\left( 0,1\right) \right\}
,L(2,1)=\left\{ \left( 1,0\right) ,\left( 1,1\right) \right\} \right\}
\end{eqnarray*}

\subsubsection{Two qubits}

The elements of $K=GF\left( 2^{2}\right) $ can be represented as 
\[
\left\{ 0,1,\lambda ,\lambda ^{2}=\lambda +1\right\} , 
\]
where $0$ is the additive identity, $1$ is the multiplicative 
identity, and $%
1+1=\lambda +\lambda =0$. The other relations follow in the obvious 
way,
such as $\lambda \left( \lambda +1\right) =\lambda ^{2}+\lambda 
=\lambda
+1+\lambda =1$. The $20$ lines of $V_{2}\left( GF\left( 4\right) 
\right) $
fall into five classes of four parallel lines each. The class of 
vertical
lines is generated by $L(4,0)=\left\{ \left( 0,0\right) ,\left( 
0,1\right)
,\left( 0,\lambda \right) ,\left( 0,\lambda +1\right) \right\} $, and 
shifts
of $L(4,0)$ by $\gamma \left( 1,0\right) .$ The other four classes are
generated by $L(\alpha ,0)=\left\{ \beta u_{\alpha }:\beta \in 
GF\left(
4\right) \right\} $ and shifts by $\gamma \left( 0,1\right) $, where 
$%
L(0,\gamma )$ corresponds to a horizontal line. Graphs of lines in $%
V_{2}\left( 2^{2}\right) $ appear in both \cite{Leonhardt} and \cite
{WoottersMUB}.

\subsection{\label{positivity}Positivity relation}

We include this brief discussion in order to illustrate the 
difficulty 
in determining whether a given phase space function corresponds to a 
positive operator. The method given here is closely related to the 
proof given in \cite
{Narcowich}. Let $\{c_{jk}\}$ be an arbitrary set of complex 
coefficients
and define the matrix $B=\sum_{j,k}c_{jk}S_{j,k}.$ Then $\rho $ $\geq 
0$ if
and only if $tr(\rho BB^{\dagger })\geq 0$ for all $B.$ Writing out 
the sum
and using the properties of the spin matrices gives 
\[
tr(\rho BB^{\dagger })=\sum_{j,k,s,t}c_{jk}c_{st}^{*}tr(\rho
S_{j-s,k-t})\eta ^{s(t-k)}. 
\]
Now we can express the trace in terms of the characteristic function 
\[
tr(\rho S_{x,y})=\left\{ 
\begin{array}{lll}
\chi (yu_{p}) & \text{if} & x=0 \\ 
\chi (xu_{a}) & \text{where} & a=x^{-1}y,\text{ }x\neq 0
\end{array}
\right. . 
\]
Therefore, we have, a not very illuminating, necessary and sufficient
condition for $\chi $ to arise from a positive matrix. The necessary 
and
sufficient condition for $\chi $ to correspond to a density matrix 
also requires
that $\chi (0)=tr\rho =1.$


\begin{thebibliography}{99}
    
 \bibitem{BBRB}  S. Bandyopadhyay, P. O. Boykin, V. Roychowdhury, F. 
Vatan, Algorithmica \textbf{34}, 512 (2002), quant-ph/0103162 (Sept. 2001).

\bibitem{Paz}  P. Bianucci, C. Miquel, P. Paz, and M. Saraceno,
quant-ph/0106091, (Jan. 2001).

\bibitem{Bell}  J.S. Bell, Physics \textbf{1},195 (1964). Reprinted 
in J. S. Bell, \textit{Speakable and unspeakable in quantum mechanics}, 
Cambridge University Press, Cambridge, 1987.  

\bibitem{Calderbank}  A. R. Calderbank, P. J Cameron, W. M. Kantor, 
J. J. Seidel, Proc. London Math. Soc.\textbf{3}, 436-480 (1997).

\bibitem{Fivel}  D. Fivel, Phys. Rev. Lett. \textbf{74}, 835 (1995).


\bibitem{Folland}  G. B. Folland, \textit{Harmonic Analysis in Phase 
Space}, Princeton University Press, Princeton, NJ, 1989, W. P. Schleich, 
\textit{Quantum Optics in Phase Space}, (Wiley-VCH, Berlin, 2001).

\bibitem{Galvao}  E. F. Galvao, quant-ph/0405070, (May 2004).

\bibitem{Woottersphase}  K. S. Gibbons, M. J. Hoffman,W. K. Wootters,
quant-ph/040115 v3 (2004).

\bibitem{review}  M. Hillery, R. F. O'Connell, M.O. Scully, and E. P.
Wigner, Phys. Rep. \textbf{106}, 121 (1984).

\bibitem{Hua}  L. K. Hua, \textit{Introduction to Number Theory}, 
(Springer, New York,1982), p. 364.

\bibitem{Ivanovic}  I. D. Ivanovi\'{c}, J. Phys. A \textbf{14}, 3241 
(1981).

\bibitem{Kon}  M. Koniorczyk, V. Buzek, and J. Hanszky, Phys. Rev. A 
64, 034301 (2001).

\bibitem{Leonhardt}  U. Leonhardt, Phys. Rev. A\textbf{\ 53}, 2998 
(1996).

\bibitem{LN}  R. Lidl and H. Niederreiter, \textit{Finite Fields},
Encyclopedia of Mathematics and its Applications Volume 20, 
Addison-Wesley, Reading MA, (1983).

\bibitem{LR}  Longenacker and Roettler, quant-ph/0309120 (Sept. 2003).

\bibitem{Narcowich}  F. J. Narcowich and R. F. O'Connell, Phys. Rev. 
A \textbf{34}, 1\ (1986).

\bibitem{Paz0} Juan Pablo Paz, Augusto Jose Roncaglia, Marcos 
Saraceno, quant-ph/0410117, (Oct. 2004).

\bibitem{Peres}  A. Peres, Phys. Rev. Lett. \textbf{77}, 1413 (1996).

\bibitem{PRfourier}  A. O. Pittenger and M. H. Rubin, Phys. Rev. A 
\textbf{62}, 032313, (2000).

\bibitem{PRMUB}  A. O. Pittenger and M. H. Rubin, Linear Alg. Appl. 
\textbf{390}, 255 (2004), quant-ph/0308142,  (Aug. 2003).


\bibitem{Schwinger}  J. Schwinger, Proc. Nat. Acad. Sci. \textbf{46}, 
570 (1960).

\bibitem{Terras}  A. Terras, \textit{Fourier Analysis on Finite 
Groups and Applications}, Cambridge University Press, Cambridge, 1999.

\bibitem{Vaccaro}  J. A. Vaccaro and D. T. Pegg, Phys. Rev. 
A\textbf{41}, 5156 (1990).


\bibitem{Vourdas}  A. Vourdas, Rep. Prog. Phys. \textbf{67}, 267 
(2004).

\bibitem{Weyl}  H. Weyl, \textit{The theory of groups and quantum 
mechanics}, (Dover Publications, New York, 1950).

\bibitem{Wig32}  E. P. Wigner, Phys. Rev. \textbf{40}, 749 (1932).

\bibitem{WoottersS}  W. K. Wootters, Found. Phys. \textbf{14}, 391 
(1986).


\bibitem{Wootters0}  W. K. Wootters, Ann. Phys. \textbf{176}, 1 
(1987).

\bibitem{WoottersMUB}  W. K. Wootters and B. D. Fields, Annals of 
Physics \textbf{191}, 363-381 (1989).


\end{thebibliography}
\end{document}